\documentclass[journal]{IEEEtran}
\ifCLASSINFOpdf
   \usepackage[pdftex]{graphicx}
\else
   \usepackage[dvips]{graphicx}
\fi
\usepackage[cmex10]{amsmath}
\usepackage{url}

\usepackage{siunitx}
\usepackage{lipsum}
\usepackage{amsbsy}
\usepackage{amsmath}
\usepackage{fixmath}
\usepackage{amssymb}
\usepackage{epsfig}
\usepackage{epstopdf}
\usepackage{algorithm,algpseudocode}

\usepackage{subfigure}
\usepackage{tikz}
\usetikzlibrary{arrows,decorations,backgrounds,shadows,plotmarks, positioning, calc,shapes, patterns,decorations.pathreplacing,chains}
\usepackage[american,cuteinductors,smartlabels]{circuitikz}
\usepackage{cite}
\usepackage{pgfplots}
\usepackage{import} 
\usepackage{float}
\usepackage{nomencl}
\makenomenclature
\pgfplotsset{compat=newest}
\pgfplotsset{plot coordinates/math parser=false}
\pgfplotsset{every axis/.append style={font=\footnotesize}}
\pgfplotsset{
    ylabel right/.style={
        after end axis/.append code={
            \node [rotate=90, anchor=north] at (rel axis cs:1,0.5) {#1};
        }   
    }
}
\newlength\figureheight
\newlength\figurewidth
\newlength\subgraphheight
\newlength\subgraphwidth
\makeatletter

\pgfkeys{
  /tikz/blockm/.initial=1,
  /tikz/blockn/.initial=1
}

\pgfdeclareshape{dspantenna}{

	\inheritsavedanchors[from=rectangle]

    \inheritanchorborder[from=rectangle]

    \inheritanchor[from=rectangle]{center}
    \inheritanchor[from=rectangle]{south}
    \inheritanchor[from=rectangle]{west}
    \inheritanchor[from=rectangle]{north}
    \inheritanchor[from=rectangle]{east}
    \inheritanchor[from=rectangle]{south east}
    \inheritanchor[from=rectangle]{south west}
    \inheritanchor[from=rectangle]{north east}
    \inheritanchor[from=rectangle]{north west}
	\saveddimen{\halfheight}{
	\pgfmathsetlength{\pgf@xa}{\pgfkeysvalueof{/pgf/minimum height}/2}
	\pgfmathsetlength{\pgf@xb}{\pgfkeysvalueof{/pgf/outer ysep}}
	\advance\pgf@xa by \pgf@xb
	\pgf@x=\pgf@xa
	}
	\saveddimen{\halfwidth}{
		\pgfmathsetlength{\pgf@xa}{\pgfkeysvalueof{/pgf/minimum width}/2}
		\pgfmathsetlength{\pgf@xb}{\pgfkeysvalueof{/pgf/outer xsep}}
		\advance\pgf@xa by \pgf@xb
		\pgf@x=\pgf@xa
		}
		
    \backgroundpath{

        \northeast \pgf@xa=\pgf@x \pgf@ya=\pgf@y
                \pgf@xb=\pgf@xa \pgf@yb=\pgf@ya
        
        \pgfpathmoveto{\pgfpoint{\pgf@xb}{\pgf@yb}}
        \advance \pgf@xb by -\halfwidth
        \advance \pgf@yb by -\halfwidth
        \pgfpathlineto{\pgfpoint{\pgf@xb}{\pgf@yb}}
        \advance \pgf@xb by -\halfwidth
        \advance \pgf@yb by \halfwidth
        \pgfpathlineto{\pgfpoint{\pgf@xb}{\pgf@yb}}
        \advance \pgf@xb by \halfwidth
        \advance \pgf@yb by -\halfwidth
        \pgfpathmoveto{\pgfpoint{\pgf@xb}{\pgf@yb}}
        \advance \pgf@yb by -\halfwidth
        \pgfpathlineto{\pgfpoint{\pgf@xb}{\pgf@yb}}
        \advance \pgf@xb by \halfwidth
        \pgfpathlineto{\pgfpoint{\pgf@xb}{\pgf@yb}}
        
        \pgfusepath{draw}
    }
}

\pgfdeclareshape{dsphdots}{

	\inheritsavedanchors[from=rectangle]

    \inheritanchorborder[from=rectangle]

    \inheritanchor[from=rectangle]{center}
    \inheritanchor[from=rectangle]{south}
    \inheritanchor[from=rectangle]{west}
    \inheritanchor[from=rectangle]{north}
    \inheritanchor[from=rectangle]{east}
    \inheritanchor[from=rectangle]{south east}
    \inheritanchor[from=rectangle]{south west}
    \inheritanchor[from=rectangle]{north east}
    \inheritanchor[from=rectangle]{north west}
	\saveddimen{\halfheight}{
	\pgfmathsetlength{\pgf@xa}{\pgfkeysvalueof{/pgf/minimum height}/2}
	\pgfmathsetlength{\pgf@xb}{\pgfkeysvalueof{/pgf/outer ysep}}
	\advance\pgf@xa by \pgf@xb
	\pgf@x=\pgf@xa
	}
	\saveddimen{\halfwidth}{
		\pgfmathsetlength{\pgf@xa}{\pgfkeysvalueof{/pgf/minimum width}/2}
		\pgfmathsetlength{\pgf@xb}{\pgfkeysvalueof{/pgf/outer xsep}}
		\advance\pgf@xa by \pgf@xb
		\pgf@x=\pgf@xa
		}
		
    \backgroundpath{

        \northeast \pgfutil@tempdima=\pgf@x \pgfutil@tempdimb=\pgf@y
		
		\pgf@xa=\halfwidth
		\pgf@ya=\halfheight
		\advance\pgfutil@tempdima by -\pgf@xa
		\advance\pgfutil@tempdimb by -\pgf@ya 
		
       \pgfpathcircle{\pgfpoint{\pgfutil@tempdima}{\pgfutil@tempdimb}}{\halfheight/3}
        \pgf@xa=\halfwidth
        \pgf@ya=\halfheight
       \advance\pgfutil@tempdima by \pgf@ya
       \advance\pgfutil@tempdima by -\pgf@xa
        \pgfpathcircle{\pgfpoint{\pgfutil@tempdima}{\pgfutil@tempdimb}}{\halfheight/3}
        \pgf@xa=\halfwidth
        \pgf@ya=\halfheight
        \advance\pgfutil@tempdima by 2\pgf@xa
        \advance\pgfutil@tempdima by -2\pgf@ya
        \pgfpathcircle{\pgfpoint{\pgfutil@tempdima}{\pgfutil@tempdimb}}{\halfheight/3} 
        
        \pgfusepath{fill}
    }
}

\pgfdeclareshape{dspvdots}{

	\inheritsavedanchors[from=rectangle]

    \inheritanchorborder[from=rectangle]

    \inheritanchor[from=rectangle]{center}
    \inheritanchor[from=rectangle]{south}
    \inheritanchor[from=rectangle]{west}
    \inheritanchor[from=rectangle]{north}
    \inheritanchor[from=rectangle]{east}
    \inheritanchor[from=rectangle]{south east}
    \inheritanchor[from=rectangle]{south west}
    \inheritanchor[from=rectangle]{north east}
    \inheritanchor[from=rectangle]{north west}
	\saveddimen{\halfheight}{
	\pgfmathsetlength{\pgf@xa}{\pgfkeysvalueof{/pgf/minimum height}/2}
	\pgfmathsetlength{\pgf@xb}{\pgfkeysvalueof{/pgf/outer ysep}}
	\advance\pgf@xa by \pgf@xb
	\pgf@x=\pgf@xa
	}
	\saveddimen{\halfwidth}{
		\pgfmathsetlength{\pgf@xa}{\pgfkeysvalueof{/pgf/minimum width}/2}
		\pgfmathsetlength{\pgf@xb}{\pgfkeysvalueof{/pgf/outer xsep}}
		\advance\pgf@xa by \pgf@xb
		\pgf@x=\pgf@xa
		}
		
    \backgroundpath{

        \northeast \pgfutil@tempdima=\pgf@x \pgfutil@tempdimb=\pgf@y
		
		\pgf@xa=\halfwidth
		\pgf@ya=\halfheight
		\advance\pgfutil@tempdima by -\pgf@xa
		\advance\pgfutil@tempdimb by -\pgf@ya 
		
       \pgfpathcircle{\pgfpoint{\pgfutil@tempdima}{\pgfutil@tempdimb}}{\halfwidth/3}
        \pgf@xa=\halfwidth
        \pgf@ya=\halfheight
       \advance\pgfutil@tempdimb by \pgf@xa
       \advance\pgfutil@tempdimb by -\pgf@ya
        \pgfpathcircle{\pgfpoint{\pgfutil@tempdima}{\pgfutil@tempdimb}}{\halfwidth/3}
        \pgf@xa=\halfwidth
        \pgf@ya=\halfheight
        \advance\pgfutil@tempdimb by 2\pgf@ya
        \advance\pgfutil@tempdimb by -2\pgf@xa
        \pgfpathcircle{\pgfpoint{\pgfutil@tempdima}{\pgfutil@tempdimb}}{\halfwidth/3} 
        
        \pgfusepath{fill}
    }
}

\pgfdeclareshape{dspmixer}{
\inheritsavedanchors[from=circle]
    \inheritanchorborder[from=circle]

    \inheritanchor[from=circle]{center}
    \inheritanchor[from=circle]{south}
    \inheritanchor[from=circle]{west}
    \inheritanchor[from=circle]{north}
    \inheritanchor[from=circle]{east}
    \inheritanchor[from=circle]{south east}
    \inheritanchor[from=circle]{south west}
    \inheritanchor[from=circle]{north east}
    \inheritanchor[from=circle]{north west}

    \inheritbackgroundpath[from=circle]

    \beforebackgroundpath{

        \radius \pgf@xb=
        0.7071067812\pgf@x
        \centerpoint \pgf@xa=\pgf@x \pgf@ya=\pgf@y \pgf@xc=\pgf@x \pgf@yc=\pgf@y

        \advance\pgf@xa by -\pgf@xb
        \advance\pgf@ya by -\pgf@xb
        \advance\pgf@xc by \pgf@xb
        \advance\pgf@yc by \pgf@xb
        \pgfpathmoveto{\pgfpoint{\pgf@xa}{\pgf@ya}}
        \pgfpathlineto{\pgfpoint{\pgf@xc}{\pgf@yc}}

        \advance\pgf@ya by 2\pgf@xb
        \advance\pgf@yc by -2\pgf@xb
        \pgfpathmoveto{\pgfpoint{\pgf@xa}{\pgf@ya}}
        \pgfpathlineto{\pgfpoint{\pgf@xc}{\pgf@yc}}

        \pgfusepath{draw}
    }
}

\pgfdeclareshape{dspadder}{
\inheritsavedanchors[from=circle]
    \inheritanchorborder[from=circle]

    \inheritanchor[from=circle]{center}
    \inheritanchor[from=circle]{south}
    \inheritanchor[from=circle]{west}
    \inheritanchor[from=circle]{north}
    \inheritanchor[from=circle]{east}
    \inheritanchor[from=circle]{south east}
    \inheritanchor[from=circle]{south west}
    \inheritanchor[from=circle]{north east}
    \inheritanchor[from=circle]{north west}

    \inheritbackgroundpath[from=circle]

    \beforebackgroundpath{

                \radius \pgf@xb=\pgf@x
        \centerpoint \pgf@xa=\pgf@x \pgf@ya=\pgf@y \pgf@xc=\pgf@x \pgf@yc=\pgf@y

        \advance\pgf@yc by \pgf@xb
        \pgfpathmoveto{\pgfpoint{\pgf@xc}{\pgf@yc}}
        \advance\pgf@yc by -2\pgf@xb
        \pgfpathlineto{\pgfpoint{\pgf@xc}{\pgf@yc}}
        \advance\pgf@xc by -\pgf@xb
        \advance\pgf@yc by \pgf@xb
        \pgfpathmoveto{\pgfpoint{\pgf@xc}{\pgf@yc}}
        \advance\pgf@xc by 2\pgf@xb
        \pgfpathlineto{\pgfpoint{\pgf@xc}{\pgf@yc}}

        \pgfusepath{draw}
    }
}

\pgfdeclareshape{dspblockmton}{

\inheritsavedanchors[from=rectangle]
    \inheritanchorborder[from=rectangle]

    \inheritanchor[from=rectangle]{center}
    \inheritanchor[from=rectangle]{south}
    \inheritanchor[from=rectangle]{west}
    \inheritanchor[from=rectangle]{north}
    \inheritanchor[from=rectangle]{east}
    \inheritanchor[from=rectangle]{south east}
    \inheritanchor[from=rectangle]{south west}
    \inheritanchor[from=rectangle]{north east}
    \inheritanchor[from=rectangle]{north west}
    \saveddimen\halfwidth{%
        \pgfmathsetlength{\pgf@xb}{\pgfkeysvalueof{/pgf/minimum width}}%
        \pgf@x=0.5\pgf@xb
      }
      \saveddimen\halfheight{
        \pgfmathsetlength{\pgf@yb}{\pgfkeysvalueof{/pgf/minimum height}}%
        \pgf@x=0.5\pgf@yb
      }
    \savedmacro\blockm{%
        \pgfmathparse{int(\pgfkeysvalueof{/tikz/blockm})}%
        \let\blockm\pgfmathresult
      }
      \savedmacro\blockn{%
          \pgfmathparse{int(\pgfkeysvalueof{/tikz/blockn})}%
          \let\blockn\pgfmathresult
        }
      
        \savedanchor\centerpoint{\pgfpointorigin}
        \anchor{center}{\centerpoint}
        \backgroundpath{%
          \pgfpathrectanglecorners{\pgfqpoint{-\halfwidth}{-\halfheight}}%
            {\pgfqpoint{\halfwidth}{\halfheight}}
        }  
        
    \pgfutil@g@addto@macro\pgf@sh@s@dspblockmton{%
        \pgfmathloop%
        \ifnum\pgfmathcounter>\blockm\relax%
        \else%
          \pgfutil@ifundefined{pgf@anchor@dspblockmton@m\pgfmathcounter}{%
            \expandafter\xdef\csname pgf@anchor@dspblockmton@m\pgfmathcounter\endcsname{%
              \noexpand\pgf@sh@lib@dspblockmton@manchor{\pgfmathcounter}%
            }%
          }{}%
        \repeatpgfmathloop%
        
      }
      
      \pgfutil@g@addto@macro\pgf@sh@s@dspblockmton{%
              
              \pgfmathloop%
                      \ifnum\pgfmathcounter>\blockn\relax%
                      \else%
                        \pgfutil@ifundefined{pgf@anchor@dspblockmton@n\pgfmathcounter}{%
                          \expandafter\xdef\csname pgf@anchor@dspblockmton@n\pgfmathcounter\endcsname{%
                            \noexpand\pgf@sh@lib@dspblockmton@nanchor{\pgfmathcounter}%
                          }%
                        }{}%
                      \repeatpgfmathloop%
            }

}

	\def\pgf@sh@lib@dspblockmton@manchor#1{%
	  \ifnum#1>\blockm\relax%
	    \pgfpointorigin%
	  \else
	    \pgfpoint{-\halfwidth}{\halfheight*2*(\blockm-2*#1+1)/(2*\blockm-1)}%
	  \fi  
	}
		\def\pgf@sh@lib@dspblockmton@nanchor#1{%
		  \ifnum#1>\blockn\relax%
		    \pgfpointorigin%
		  \else
		    \pgfpoint{\halfwidth}{\halfheight*2*(\blockn-2*#1+1)/(2*\blockn-1)}%
		  \fi  
		}
\pgfdeclareshape{dsptriangler}{

	\inheritsavedanchors[from=rectangle]

    \inheritanchorborder[from=rectangle]

    \inheritanchor[from=rectangle]{center}
    \inheritanchor[from=rectangle]{south}
    \inheritanchor[from=rectangle]{west}
    \inheritanchor[from=rectangle]{north}
    \inheritanchor[from=rectangle]{east}
    \inheritanchor[from=rectangle]{south east}
    \inheritanchor[from=rectangle]{south west}
    \inheritanchor[from=rectangle]{north east}
    \inheritanchor[from=rectangle]{north west}
	\saveddimen{\halfheight}{
	\pgfmathsetlength{\pgf@xa}{\pgfkeysvalueof{/pgf/minimum height}/2}
	\pgfmathsetlength{\pgf@xb}{\pgfkeysvalueof{/pgf/outer ysep}}
	\advance\pgf@xa by \pgf@xb
	\pgf@x=\pgf@xa
	}
	\saveddimen{\halfwidth}{
		\pgfmathsetlength{\pgf@xa}{\pgfkeysvalueof{/pgf/minimum width}/2}
		\pgfmathsetlength{\pgf@xb}{\pgfkeysvalueof{/pgf/outer xsep}}
		\advance\pgf@xa by \pgf@xb
		\pgf@x=\pgf@xa
		}
		
    \backgroundpath{

        \pgfutil@tempdima=\halfheight
        \pgfutil@tempdimb=\halfwidth
        \southwest \pgf@xa=\pgf@x \pgf@ya=\pgf@y 
		\pgfpathmoveto{\pgfpoint{\pgf@xa}{\pgf@ya}}
		\advance \pgf@ya by 2\pgfutil@tempdima
     	\pgfpathlineto{\pgfpoint{\pgf@xa}{\pgf@ya}}
     	\advance \pgf@xa by 2\pgfutil@tempdimb
     	\advance \pgf@ya by -\pgfutil@tempdima
     	\pgfpathlineto{\pgfpoint{\pgf@xa}{\pgf@ya}}	
     	\advance \pgf@xa by -2\pgfutil@tempdimb
     	\advance \pgf@ya by -\pgfutil@tempdima
     	\pgfpathlineto{\pgfpoint{\pgf@xa}{\pgf@ya}}       
        \pgfusepath{draw}
    }
}

\pgfdeclareshape{dsptrianglel}{

	\inheritsavedanchors[from=rectangle]

    \inheritanchorborder[from=rectangle]

    \inheritanchor[from=rectangle]{center}
    \inheritanchor[from=rectangle]{south}
    \inheritanchor[from=rectangle]{west}
    \inheritanchor[from=rectangle]{north}
    \inheritanchor[from=rectangle]{east}
    \inheritanchor[from=rectangle]{south east}
    \inheritanchor[from=rectangle]{south west}
    \inheritanchor[from=rectangle]{north east}
    \inheritanchor[from=rectangle]{north west}
	\saveddimen{\halfheight}{
	\pgfmathsetlength{\pgf@xa}{\pgfkeysvalueof{/pgf/minimum height}/2}
	\pgfmathsetlength{\pgf@xb}{\pgfkeysvalueof{/pgf/outer ysep}}
	\advance\pgf@xa by \pgf@xb
	\pgf@x=\pgf@xa
	}
	\saveddimen{\halfwidth}{
		\pgfmathsetlength{\pgf@xa}{\pgfkeysvalueof{/pgf/minimum width}/2}
		\pgfmathsetlength{\pgf@xb}{\pgfkeysvalueof{/pgf/outer xsep}}
		\advance\pgf@xa by \pgf@xb
		\pgf@x=\pgf@xa
		}
		
    \backgroundpath{

        \pgfutil@tempdima=\halfheight
        \pgfutil@tempdimb=\halfwidth
        \northeast \pgf@xa=\pgf@x \pgf@ya=\pgf@y 
		\pgfpathmoveto{\pgfpoint{\pgf@xa}{\pgf@ya}}
		\advance \pgf@ya by -2\pgfutil@tempdima
     	\pgfpathlineto{\pgfpoint{\pgf@xa}{\pgf@ya}}
     	\advance \pgf@xa by -2\pgfutil@tempdimb
     	\advance \pgf@ya by \pgfutil@tempdima
     	\pgfpathlineto{\pgfpoint{\pgf@xa}{\pgf@ya}}
     	\advance \pgf@xa by 2\pgfutil@tempdimb
     	\advance \pgf@ya by \pgfutil@tempdima	
     	\pgfpathlineto{\pgfpoint{\pgf@xa}{\pgf@ya}}       
        \pgfusepath{draw}
    }
}

\pgfdeclareshape{dsptriangled}{

	\inheritsavedanchors[from=rectangle]

    \inheritanchorborder[from=rectangle]

    \inheritanchor[from=rectangle]{center}
    \inheritanchor[from=rectangle]{south}
    \inheritanchor[from=rectangle]{west}
    \inheritanchor[from=rectangle]{north}
    \inheritanchor[from=rectangle]{east}
    \inheritanchor[from=rectangle]{south east}
    \inheritanchor[from=rectangle]{south west}
    \inheritanchor[from=rectangle]{north east}
    \inheritanchor[from=rectangle]{north west}
    
	\saveddimen{\halfheight}{
	\pgfmathsetlength{\pgf@xa}{\pgfkeysvalueof{/pgf/minimum height}/2}
	\pgfmathsetlength{\pgf@xb}{\pgfkeysvalueof{/pgf/outer ysep}}
	\advance\pgf@xa by \pgf@xb
	\pgf@x=\pgf@xa
	}
	\saveddimen{\halfwidth}{
		\pgfmathsetlength{\pgf@xa}{\pgfkeysvalueof{/pgf/minimum width}/2}
		\pgfmathsetlength{\pgf@xb}{\pgfkeysvalueof{/pgf/outer xsep}}
		\advance\pgf@xa by \pgf@xb
		\pgf@x=\pgf@xa
		}
	\anchor{east contact}{
	    \pgfutil@tempdima=\halfheight
	    \pgfutil@tempdimb=\halfwidth
	    \northeast \pgf@xa=\pgf@x \pgf@ya=\pgf@y
	    \advance \pgf@xa by -0.5\pgfutil@tempdimb
	    \advance \pgf@ya by -\pgfutil@tempdima
	    \pgf@x=\pgf@xa
	    \pgf@y=\pgf@ya		
	}
	\anchor{west contact}{
		    \pgfutil@tempdima=\halfheight
		    \pgfutil@tempdimb=\halfwidth
		    \southwest \pgf@xa=\pgf@x \pgf@ya=\pgf@y
		    \advance \pgf@xa by \pgfutil@tempdimb/2
		    \advance \pgf@ya by \pgfutil@tempdima
		    \pgf@x=\pgf@xa
		    \pgf@y=\pgf@ya		
		}	
    \backgroundpath{

        \pgfutil@tempdima=\halfheight
        \pgfutil@tempdimb=\halfwidth
        \northeast \pgf@xa=\pgf@x \pgf@ya=\pgf@y 
		\pgfpathmoveto{\pgfpoint{\pgf@xa}{\pgf@ya}}
		\advance \pgf@xa by -2\pgfutil@tempdima
     	\pgfpathlineto{\pgfpoint{\pgf@xa}{\pgf@ya}}
     	\advance \pgf@xa by \pgfutil@tempdimb
     	\advance \pgf@ya by -2\pgfutil@tempdima
     	\pgfpathlineto{\pgfpoint{\pgf@xa}{\pgf@ya}}
     	\advance \pgf@xa by \pgfutil@tempdimb
     	\advance \pgf@ya by 2\pgfutil@tempdima	
     	\pgfpathlineto{\pgfpoint{\pgf@xa}{\pgf@ya}}       
        \pgfusepath{draw}
    }
}

\tikzset{arc style/.initial={}}
\pgfdeclareshape{circle with arcs}{
    \inheritsavedanchors[from=circle]
    \inheritanchorborder[from=circle]

    \inheritanchor[from=circle]{center}
    \inheritanchor[from=circle]{south}
    \inheritanchor[from=circle]{west}
    \inheritanchor[from=circle]{north}
    \inheritanchor[from=circle]{east}

    \inheritbackgroundpath[from=circle]

    \beforebackgroundpath{
        \pgfkeys{/tikz/arc style/.get=\tmp}
        \expandafter\tikzset\expandafter{\tmp}
        \tikz@options

        \radius \pgf@xa=\pgf@x
        \centerpoint \pgf@xb=\pgf@x \pgf@yb=\pgf@y

        \advance\pgf@yb by\pgf@xa
        \pgfpathmoveto{\pgfpoint{\pgf@xb}{\pgf@yb}}
        \pgfpathlineto{\centerpoint}
        \pgfpatharc{180}{270}{\pgf@xa}

        \advance\pgf@yb by -2\pgf@xa
        \pgfpathmoveto{\pgfpoint{\pgf@xb}{\pgf@yb}}
        \pgfpatharc{0}{90}{\pgf@xa}

        \pgfusepath{draw}
    }
}
\makeatother

\usepackage{pgfplots}
\usepackage{pgfkeys}
\pgfplotsset{compat=newest}
\pgfplotsset{plot coordinates/math parser=false}

\DeclareMathAlphabet{\mathbit}{OML}{cmr}{bx}{it}
\DeclareMathAlphabet{\mathsf}{OT1}{cmss}{m}{n}
\DeclareMathAlphabet{\mathbsf}{OT1}{cmss}{bx}{it}

\begin{document}
%
\title{A Sphere Decoding Algorithm for Multistep Sequential Model Predictive
	Control}

\author{\IEEEauthorblockN{Ferdinand Grimm \IEEEmembership{Student Member,~IEEE,}, Zhenbin Zhang \IEEEmembership{Senior Member,~IEEE,} and Mehdi Baghdadi
}
}

\maketitle

\begin{abstract}
This paper investigates the combination of two model predictive control concepts,
sequential model predictive control and long-horizon model predictive control for power electronics.
To achieve sequential model predictive control, the optimization problem is split into two subproblems: The first one summarizes all control goals which linearly depend on the system inputs.
Sequential model predictive control generally requires to obtain more than one solution for the first subproblem.
Due to the mixed-integer nature of finite control set model predictive control power electronics a special sphere decoder is therefore proposed within the paper.
The second subproblem consists of all those control goals which depend nonlinearly on the system inputs and is solved by an exhaustive search.
	The effectiveness of the proposed method is validated via numerical simulations at different scenarios on a three-level neutral point clamped permanent magnet synchronous generator wind turbine system and compared to other long-horizon model predictive control methods. 
\end{abstract}


%
\IEEEpeerreviewmaketitle

\section{Introduction}\label{sec:intro}
\begin{figure}[!htbp]
	\begin{center}
		\hspace{-2.5ex}
\includegraphics[scale=0.3]{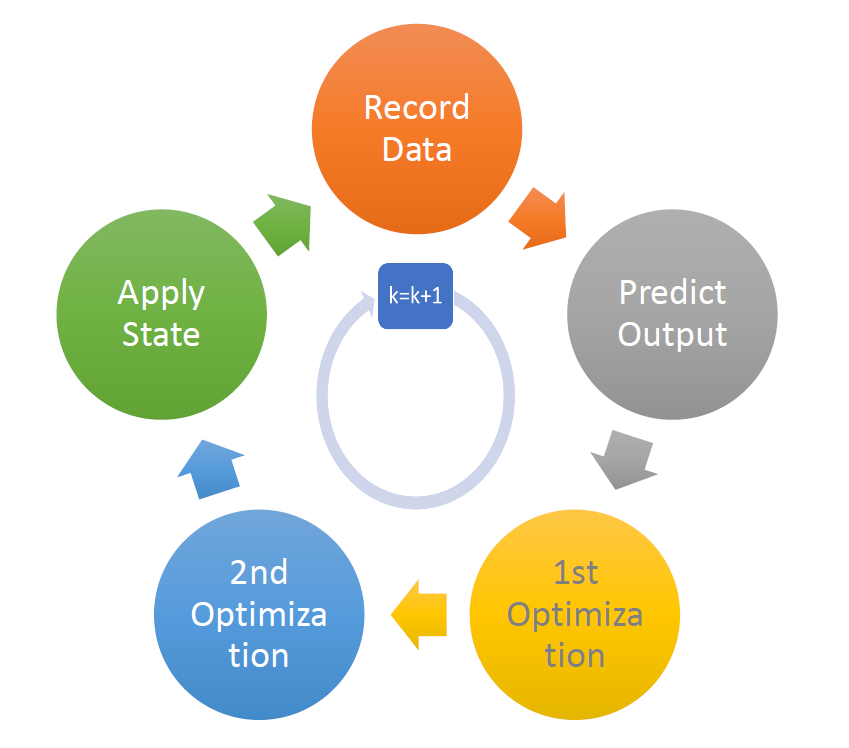}
\end{center}
	\vspace{-2ex}
	\caption{The basic principle of sequential MPC: During each sampling step the controller acquires the sensor data, predicts the output, and then solves the optimization problem corresponding to the first subsystem. The best $N_k$ candidates are then passed on to the second optimization problem. The optimal solution candidate for the second optimization problem is then applied to the system.
	This procedure is repeated at the next sampling step.}
	\label{SMPC}
\end{figure}

Model predictive control (MPC) \cite{GenMPC2} is a popular approach for the control of power electronic systems.
One of its advantages are fast dynamics \cite{GenMPC3} and the simple inclusion of different control goals as shown e.g. in \cite{GenMPC4}.
Recently, a  model predictive control approach for power electronics, sequential model predictive control
has been discussed in \cite{Seq1}. 
Predictive Power Control (PPC) is an MPC technique that aims to control both, the active as well as the reactive power which is flowing into the system. 
It has been introduced for the back-to-back converter in \cite{PPC}.
Based on PPC, several more advanced model predictive control methods have been proposed in \cite{MyThesis}.
 In \cite{b2b3} the framework was extended by using a quasi-centralized model predictive control approach. 
The performance of model predictive control can be enhanced by extending the prediction horizon \cite{LH,Multi,Multi2,Multi3}. These increases in performance however come at a price of higher computational cost \cite{LH}.
For linear systems, the computational cost can be reduced using specialized algorithms such as the sphere decoder \cite{Multi},\cite{Multi2}. 
For nonlinear systems however the applicability of efficient approximations is limited \cite{Multi3}.
\newline
One example of a popular nonlinear system in power electronics is the Three-level neutral point clamped (NPC) back-to-back PMSG wind turbine system \cite{MyThesis}. The Three-level NPC back-to-back PMSG system is a promising technology in the field of renewable energy applications \cite{Wind,PPC,MyThesis,b2b3}. A detailed description of the system is given in \cite{Wind}.
Other nonlinear control methods for the Three-level NPC back-to-back PMSG systems include Fuzzy control \cite{Gen7}, ${H_\infty }$-control \cite{Gen6} or Feedback linearization \cite{Gen5}. Alternatively, the system can be controlled by adding an additional DC/DC converter at the DC-link to control the nonlinear capacitor voltage \cite{Gen1}.
 The system consists of both, linear and nonlinear components and therefore serves as a case study system in this paper.
 \newline
Several attempts to bypass this have been made using linearization or limiting the number of states of interest \cite{Thesis}. Those concepts have been successfully applied to the  Three-level NPC back-to-back PMSG wind turbine system in \cite{FG3}.
The approximation of nonlinear cost functions allows applying the sphere decoding algorithm even though the system is nonlinear. However, those approximations are either simple or relying on many parameters to tune.
\newline
Recently \cite{Seq1,Seq2,Seq3,Seq4,Seq5,unknown} presented novel cascaded model predictive control frameworks called sequential MPC. The biggest advantage of sequential MPC is that it is not dependent on weighting factors. Instead, the two most optimal solutions for the torque are computed in a first step and from those two solutions, the one that optimizes the magnetic flux is selected, (see Fig. \ref{SMPC}).
This framework has subsequently been extended to other systems three-level NPC converters \cite{Seq4}, 
 three-level NPC wind turbine systems \cite{unknown}, and matrix converters \cite{Seq5}, as well as other modeling approaches including field weakening \cite{Seq3} and generalized predictive control \cite{Seq2}.
 \newline
Using simulations it was shown in \cite{GGG} that the current total harmonic distortion of a system can be reduced especially for low switching frequencies if a multistep MPC scheme is selected instead of a single-step MPC scheme.
Recent advantages presented in \cite{GenMPC1} showed the effectiveness of the sphere decoder with experiment results using a control period of $2$ ms for a horizon length of $5$.
 \newline
The contributions of this paper  can be summarized in the following:
\begin{enumerate}
    \item Definition of a framework that combines the two concepts of multistep model predictive control and sequential model predictive control.
    \item Formulation of a sphere decoding algorithm to efficiently solve the linear subproblem.
    \item Comparison of the proposed method with existing long-horizon model predictive control strategies on a Three-level NPC PMSG wind turbine system.
\end{enumerate}
In this paper a multistep sequential predictive control framework is proposed.
Similarly to sequential MPC, the system is distributed into several subsystems.
Therefore the system is split into a linear and nonlinear part.
Since it is necessary to utilize an efficient optimization algorithm in multistep model predictive control, a specialized algorithm is proposed for the linear part in this paper.
It is based on the sphere decoding algorithm, however, instead of only giving the best solution of a mixed-integer linear least squares problem, it computes its best $N_k$ solutions. 
First, the linear part is solved using a modified sphere decoder. The resulting $N_k$ solutions are then used to optimize the DC-link voltage balance, which is the nonlinear component of the system. Of those $N_k$ solutions, we then select the candidate that optimizes the nonlinear component using exhaustive search.
Moreover, the framework is presented with the help of a Three-Level NPC back-to-back PMSG case-study system and compared to other long-horizon MPC schemes.

 The proposed method is based on \cite{Original}, which originated this paper.
\newline
This paper is organized as follows:
In section II the system model of the sample system is given. Section III derives the resulting optimization problem. The proposed solution is derived in section IV. Section V presents the results and Section VI gives the conclusion.

\section{System Model}\label{sec:model}
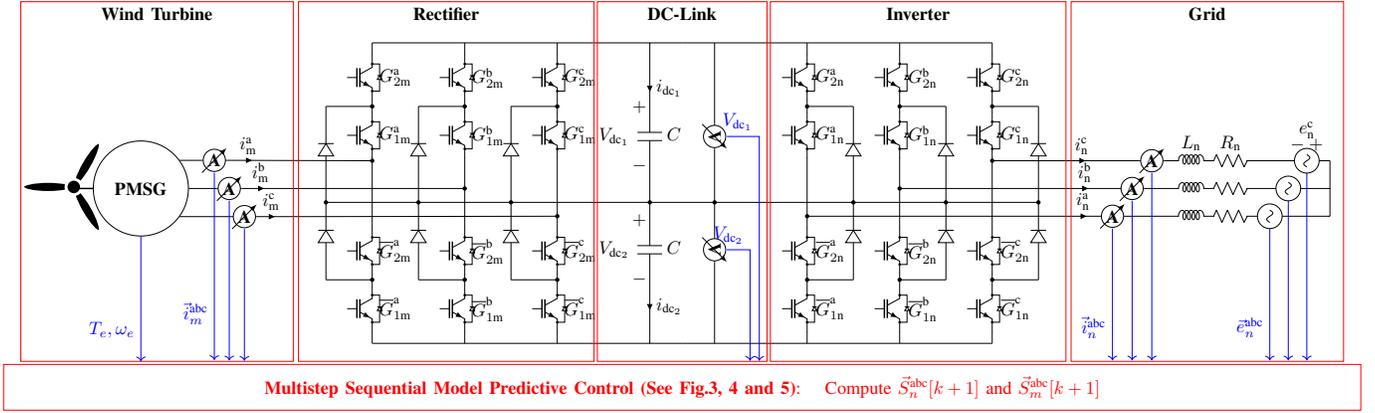
\begin{figure*}[!htbp]
	\ctikzset{bipoles/thickness=1}
	\ctikzset{bipoles/length=0.8cm}
	\ctikzset{tripoles/thyristor/height=.8}
	\ctikzset{tripoles/thyristor/width=1}
	\ctikzset{bipoles/diode/height=.375}
	\ctikzset{bipoles/diode/width=.3}
	\tikzstyle{block} = [draw,fill=white, rectangle, minimum height=1cm, minimum width=6em]
	\tikzstyle{sum} = [draw, fill=white, circle, node distance=1cm]
	\tikzstyle{pinstyle} = [pin edge={to-,thin,black}]
	\tikzset{%
		wind turbine/.pic={
			\tikzset{path/.style={fill, draw=white, ultra thick, line join=round}}
			\foreach \i in {300, 60, 180}{
				\ifcase#1
				\or
				\path [path, shift=(90:3), rotate=\i] 
				(.5,-.1075) arc (270:90:.5 and .1075) arc (90:-90:0.5 and .1075);
				\or
				\path [path, shift=(90:3), rotate=\i] 
				(0,0.125) -- (2,0.125) -- (2,0) -- (0.5,-0.375) -- cycle;
				\or
				\path [path, shift=(90:3), rotate=\i]
				(0,-0.125) arc (180:0:1 and 0.125) -- ++(0,0.125) arc (0:180:1 and 0.25) -- cycle;
				\fi
			}
			\path [path] (0,3) circle [radius=.15];
	}}
	\begin{center}
		\resizebox{\linewidth}{!}{
			\begin{tikzpicture}[scale=0.9]
			\draw
			(-6.7,0.75) node[nigbt,bodydiode](igbt1a){}
			(-6.2,0.75) node{$\overline{G}^{\textrm{a}}_{2\textrm{m}}$}
			(-6.7,-0.5) node[nigbt,bodydiode](igbt1aa){}
			(-6.2,-0.5) node{$\overline{G}^{\textrm{a}}_{1\textrm{m}}$}
			(-6.7,3.25) node[nigbt,bodydiode](igbt1b){}
			(-6.2,3.25) node{$G^{\textrm{a}}_{1\textrm{m}}$}
			(-6.7,4.5) node[nigbt,bodydiode](igbt1bb){}
			(-6.2,4.5) node{$G^{\textrm{a}}_{2\textrm{m}}$}
			(-4.7,0.75) node[nigbt,bodydiode](igbt2a){}
			(-4.2,0.75) node{$\overline{G}^{\textrm{b}}_{2\textrm{m}}$}
			(-4.7,-0.5) node[nigbt,bodydiode](igbt2aa){}
			(-4.2,-0.5) node{$\overline{G}^{\textrm{b}}_{1\textrm{m}}$}
			(-4.7,3.25) node[nigbt,bodydiode](igbt2b){}
			(-4.2,3.25) node{$G^{\textrm{b}}_{1\textrm{m}}$}
			(-4.7,4.5) node[nigbt,bodydiode](igbt2bb){}
			(-4.2,4.5) node{$G^{\textrm{b}}_{2\textrm{m}}$}
			(-2.7,0.75) node[nigbt,bodydiode](igbt3a){}
			(-2.2,0.75) node{$\overline{G}^{\textrm{c}}_{2\textrm{m}}$}
			(-2.7,-0.5) node[nigbt,bodydiode](igbt3aa){}
			(-2.2,-0.5) node{$\overline{G}^{\textrm{c}}_{1\textrm{m}}$}
			(-2.7,3.25) node[nigbt,bodydiode](igbt3b){}
			(-2.2,3.25) node{$G^{\textrm{c}}_{1\textrm{m}}$}
			(-2.7,4.5) node[nigbt,bodydiode](igbt3bb){}
			(-2.2,4.5) node{$G^{\textrm{c}}_{2\textrm{m}}$}; 
			\draw
			(6.7,0.75) node[nigbt,bodydiode](igbt6a){}
			(7.2,0.75) node{$\overline{G}^{\textrm{c}}_{2\textrm{n}}$}
			(6.7,-0.5) node[nigbt,bodydiode](igbt6aa){}
			(7.2,-0.5) node{$\overline{G}^{\textrm{c}}_{1\textrm{n}}$}
			(6.7,3.25) node[nigbt,bodydiode](igbt6b){}
			(7.2,3.25) node{$G^{\textrm{c}}_{1\textrm{n}}$}
			(6.7,4.5) node[nigbt,bodydiode](igbt6bb){}
			(7.2,4.5) node{$G^{\textrm{c}}_{2\textrm{n}}$}
			(4.7,0.75) node[nigbt,bodydiode](igbt5a){}
			(5.2,0.75) node{$\overline{G}^{\textrm{b}}_{2\textrm{n}}$}
			(4.7,-0.5) node[nigbt,bodydiode](igbt5aa){}
			(5.2,-0.5) node{$\overline{G}^{\textrm{b}}_{1\textrm{n}}$}
			(4.7,3.25) node[nigbt,bodydiode](igbt5b){}
			(5.2,3.25) node{$G^{\textrm{b}}_{1\textrm{n}}$}
			(4.7,4.5) node[nigbt,bodydiode](igbt5bb){}
			(5.2,4.5) node{$G^{\textrm{b}}_{2\textrm{n}}$}
			(2.7,0.75) node[nigbt,bodydiode](igbt4a){}
			(3.2,0.75) node{$\overline{G}^{\textrm{a}}_{2\textrm{n}}$}
			(2.7,-0.5) node[nigbt,bodydiode](igbt4aa){}
			(3.2,-0.5) node{$\overline{G}^{\textrm{a}}_{1\textrm{n}}$}
			(2.7,3.25) node[nigbt,bodydiode](igbt4b){}
			(3.2,3.25) node{$G^{\textrm{a}}_{1\textrm{n}}$}
			(2.7,4.5) node[nigbt,bodydiode](igbt4bb){}
			(3.2,4.5) node{$G^{\textrm{a}}_{2\textrm{n}}$}; 
			\draw
			(2.7,-1.25) to (igbt4aa.E)
			(igbt4aa.C)
			to (igbt4a.E)
			(igbt4a.C) to (2.7,1.5)
			-- ++(0,1)
			to (igbt4b.E) 
			(igbt4b.C) 
			to (igbt4bb.E) 
			(igbt4bb.C) to (2.7,5.25) coordinate (leg4)
			(4.7,-1.25)
			to (igbt5aa.E)
			(igbt5aa.C)
			to (igbt5a.E)
			(igbt5a.C)
			to (4.7,1.5)
			-- ++(0,1)
			to (igbt5b.E) 
			(igbt5b.C) 
			to (igbt5bb.E) 
			(igbt5bb.C) to  (4.7,5.25) coordinate (leg5)
			(6.7,-1.25)
			to (igbt6aa.E)
			(igbt6aa.C) 
			to (igbt6a.E)
			(igbt6a.C)
			to (6.7,1.5)
			-- ++(0,1)
			to (igbt6b.E) 
			(igbt6b.C) 
			to (igbt6bb.E) 
			(igbt6bb.C) to (6.7,5.25) coordinate (leg6);
			\node[circ] at (-6.7,0.125) {};
			\node[circ] at (-4.7,0.125) {};
			\node[circ] at (-2.7,0.125) {};
			\node[circ] at (2.7,0.125) {};
			\node[circ] at (4.7,0.125) {};
			\node[circ] at (6.7,0.125) {};
			\node[circ] at (-6.7,3.875) {};
			\node[circ] at (-4.7,3.875) {};
			\node[circ] at (-2.7,3.875) {};
			\node[circ] at (2.7,3.875) {};
			\node[circ] at (4.7,3.875) {};
			\node[circ] at (6.7,3.875) {};
			\draw(-6.7,0.125) -- (-7.7,0.125);
			\draw(-4.7,0.125) -- (-5.7,0.125);
			\draw(-2.7,0.125) -- (-3.7,0.125);
			\draw(6.7,0.125)  -- (7.7,0.125);
			\draw(4.7,0.125)  -- (5.7,0.125);
			\draw(2.7,0.125)  -- (3.7,0.125);
			\draw(-6.7,3.875) -- (-7.7,3.875);
			\draw(-4.7,3.875) -- (-5.7,3.875);
			\draw(-2.7,3.875) -- (-3.7,3.875);
			\draw(6.7,3.875)  -- (7.7,3.875);
			\draw(4.7,3.875)  -- (5.7,3.875);
			\draw(2.7,3.875)  -- (3.7,3.875);
			\draw(-7.7,0.125) to[diode]  (-7.7,2);
			\draw(-5.7,0.125) to[diode]  (-5.7,2);
			\draw(-3.7,0.125) to[diode]  (-3.7,2);
			\draw(7.7,0.125) to[diode]  (7.7,2);
			\draw(5.7,0.125) to[diode]  (5.7,2);
			\draw(3.7,0.125) to[diode]  (3.7,2);
			\draw(-7.7,2) to[diode]  (-7.7,3.875);
			\draw(-5.7,2) to[diode]  (-5.7,3.875);
			\draw(-3.7,2) to[diode]  (-3.7,3.875);
			\draw(7.7,2) to[diode]  (7.7,3.875);
			\draw(5.7,2) to[diode]  (5.7,3.875);
			\draw(3.7,2) to[diode]  (3.7,3.875);
			\draw 
		    (6.7,5.25) 	
			-- ++(-6.7,0)
			(0,-1.25)--(6.7,-1.25)
			;
			\draw
			(-2.7,-1.25)
			to (igbt3aa.E)
			(igbt3aa.C) 
			to (igbt3a.E)
			(igbt3a.C)			
			to (-2.7,1.5)
			-- ++(0,1)
			to (igbt3b.E) 
			(igbt3b.C) 
			to (igbt3bb.E) 
			(igbt3bb.C) to (-2.7,5.25) coordinate (leg3)
			(-4.7,-1.25)
			to (igbt2aa.E)
			(igbt2aa.C) 
			to (igbt2a.E)
			(igbt2a.C)
			to (-4.7,1.5)
			-- ++(0,1)
			to (igbt2b.E) 
			(igbt2b.C) 
			to (igbt2bb.E) 
			(igbt2bb.C) to (-4.7,5.25) coordinate (leg2)
			(-6.7,-1.25)
			to (igbt1aa.E)
			(igbt1aa.C) 
			to (igbt1a.E)
			(igbt1a.C)
			to (-6.7,1.5)
			-- ++(0,1)
			to (igbt1b.E) 
			(igbt1b.C) 
			to (igbt1bb.E) 
			(igbt1bb.C)  to (-6.7,5.25) coordinate (leg1);
			\draw 
			(-6.7,5.25)
			-- ++(6,0)
			-- ++(0,-0.8)
			to[C, l=$C$, v=$V_{\textrm{dc}_1}$, i=$i_{\textrm{dc}_1}$, current/distance=-1.0]++(0,-2.45)
			to[C, l=$C$, v=$V_{\textrm{dc}_2}$, i=$i_{\textrm{dc}_2}$, current/distance=1.0]  ++(0,-2.45)
			|- (-6.7,-1.25);
			\draw
			(0.7,5.25)--++(0,-0.8) 
			to[voltmeter,n=VDC1] ++ (0,-2.45)
			to[voltmeter,n=VDC2] ++ (0,-2.45)
			-- (0.7,-1.25);
			\draw
			(-0.7,5.25) -- (0.0,5.25)
			(-0.7,-1.25) -- (0.0,-1.25);
			\node[circ] at (0.7,1.8) {};
			\node[circ] at (-0.7,1.8) {};
			\node[circ] at (3.7,1.8) {};
			\node[circ] at (5.7,1.8) {};
			\node[circ] at (7.7,1.8) {};
			\node[circ] at (-3.7,1.8) {};
			\node[circ] at (-5.7,1.8) {};
			\node[circ] at (-7.7,1.8) {};
			\draw (-0.7,1.8) -- (0.7,1.8);
			\draw (-7.7,1.8) -- (-0.7,1.8);
			\draw (7.7,1.8) -- (0.7,1.8);
			\node[circle,draw, inner sep=1pt,minimum size=1.85cm] (PMSG) at (-11.7,2.1) {\textbf{PMSG}};
			\draw[white]
			(-13.15,1.5) coordinate (V3)
			to ++(0,0.6) coordinate (V2)
			(V2) to ++(0,0.6) coordinate (V1);
			\draw
			(PMSG.0) to ++(0.1,0) to[ammeter, i^>=$i_{\textrm{m}}^{\textrm{b}}$, current/distance=0.8,n=Ampm2] ++ (1.55,0)   to (leg2 |- V2) node [circ] {}
			(PMSG.36) to ++(0.25,0)  to[ammeter, i^>=$i_{\textrm{m}}^{\textrm{a}}$, current/distance=2.0,n=Ampm1] ++ (1.0,0)   to (leg1 |- V1) node [circ] {}
			(PMSG.324) to ++(0.1,0) to[ammeter, i^>=$i_{\textrm{m}}^{\textrm{c}}$, current/distance=0.3,n=Ampm3] ++ (2.6,0)   to (leg3 |- V3) node [circ] {};
			\draw
			(-13.15,-1.2) pic {wind turbine=1};
			\draw[thick]
			(V2)--(PMSG); 
			\draw
			(14.0,1.5) coordinate (V4)
			to ++(0,0.6) coordinate (V5)
			(V5) to ++(0,0.6) coordinate (V6)
			(V5) to ++(-0.4,0) to[sV,n=V5] ++ (-1.0,0) to ++(-0.4,0) to[R] ++ (-0.7,0) to[L,mirror] ++(-1.0,0) to[ammeter, i_<=$i_{\textrm{n}}^{\textrm{b}}$, current/distance=1.5,mirror,invert,n=Ampm5] ++ (-1.55,0)   to (leg5 |- V5) node [circ] {}
			(V6) to[sV_=$e_{\textrm{n}}^{\textrm{c}}$,n=V6] ++ (-1.0,0) to ++(-0.8,0) to[R,a=$R_{\textrm{n}}$] ++ (-0.7,0) to[L,a=$L_{\textrm{n}}$,mirror]++(-1.0,0) to[ammeter, i_<=$i_{\textrm{n}}^{\textrm{c}}$, current/distance=15.0,mirror,invert,n=Ampm6] ++ (-0.7,0)   to (leg6 |- V6) node [circ] {}
			(V4)to ++(-0.8,0) to[sV,n=V4] ++ (-1,0) to[R] ++ (-0.7,0) to[L,mirror] ++(-1.0,0) to[ammeter, i_<=$i_{\textrm{n}}^{\textrm{a}}$, current/distance=0.4,mirror,invert,n=Ampm4] ++ (-2.4,0)   to (leg4 |- V4) node [circ] {};
			\draw[blue,->]	(PMSG.south) -- ++ (0,-2.7) node[left,near end] {$T_e, \omega_e$};
			\draw[blue,->]
			(V4.north) -- ++(0,-2.85) node[left,near end] {$\vec{e}_n^{\textrm{abc}}$};
		    \draw[blue,->]	(V5.north) -- ++(0,-3.45);
			\draw[blue,->] (V6.north) -- ++(0,-4.05);
			\draw[blue,->]
			(Ampm1.south) -- ++(0,-4.05) node[left,near end] {$\vec{i}_m^{\textrm{abc}}$};
			\draw[blue,->](Ampm2.south) -- ++(0,-3.45);
			\draw[blue,->](Ampm3.south) -- ++(0,-2.85);
			\draw[blue,->](Ampm4.south) -- ++(0,-2.85) node[left,near end] {$\vec{i}_n^{\textrm{abc}}$};
			\draw[blue,->](Ampm5.south) -- ++(0,-3.45); 
			\draw[blue,->](Ampm6.south) -- ++(0,-4.05);
			\draw[blue,->]
			(VDC1.north) -| ++(0.7,-4.85) node[above left,midway] {$V_{\textrm{dc}_1}$};
			\draw[blue,->]
			(VDC2.north) -| ++(0.5,-2.4) node[above left,midway] {$V_{\textrm{dc}_2}$};
			\node [
			rectangle,draw,red,
			minimum width=3.3cm,
			minimum height=7.0cm,
			] (DC-Box) at (0,2.25) {};
			\node[below] at (DC-Box.north) {\textbf{DC-Link}};
			\node [
			rectangle,draw,red,
			minimum width=5.75cm,
			minimum height=7.0cm,
			] (Rect-Box) at (-5.1,2.25) {};
			\node[below] at (Rect-Box.north) {\textbf{Rectifier}};
			\node [
			rectangle,draw,red,
			minimum width=5.75cm,
			minimum height=7.0cm,
			] (Inv-Box) at (5.1,2.25) {};
			\node[below] at (Inv-Box.north) {\textbf{Inverter}};
			\node [
			rectangle,draw,red,
			minimum width=5.3cm,
			minimum height=7.0cm,
			] (Line-Box) at (-11.35,2.25) {};
			\node[below] at (Line-Box.north) {\textbf{Wind Turbine}};
			\node [
			rectangle,draw,red,
			minimum width=5.3cm,
			minimum height=7.0cm,
			] (Load-Box) at (11.35,2.25) {};
			\node[below] at (Load-Box.north) {\textbf{Grid}};
			\node [
			rectangle,draw,red,
			minimum width=26.4cm,
			minimum height=0.89cm,
			] (Controller-Box) at (0,-2.20) {\textbf{Multistep Sequential Model Predictive Control (See Fig.3, 4 and 5)}: \quad Compute $\vec{S}_n^{\textrm{abc}}[k+1]$ and $\vec{S}_m^{\textrm{abc}}[k+1]$};
			\node[below] at (Controller-Box.north) {};
			\end{tikzpicture}}
	\end{center}
	\vspace{-2ex}
	\caption{Block diagram of the Three-level back-to-back PMSG wind turbine system consisting of the wind turbine, the grid side with RL-filter, and the converter\cite{MyThesis} as well as the proposed control scheme. The parameters are explained in Table \ref{tab:mac}.}
	\label{fig:sys}
\end{figure*}

The proposed system considered in this paper consists of a Three-level back-to-back PMSG wind turbine systems with RL-filter. An overview of the system is shown in Fig. \ref{fig:sys}.

\subsection{Converter with DC-Link}
The back-to-back converter system considered in this paper consists of two sides: The machine side which is connected to the rectifier part of the converter and the grid side which is connected to the inverter part of the converter \cite{MyThesis}.
Both sides are connected to each other through the DC-link.
The converter takes a certain switching state in abc-coordinates $\vec{S}_i^{\textrm{abc}} \in \{-1,0,1\}^3$ for $i \in \{m,n\}$ as input and produces a three-phase voltage $\vec{u}_i^{\textrm{abc}}$ on the respective side as an output.
The a, b, c coordinates represent the Three-phase $a,b,c$- wires in Fig. \ref{fig:sys}.
The relationship between the  switching state $\vec{S}_i^{\textrm{abc}}$ and the voltage $\vec{u}_i^{\textrm{abc}}$ is given by
\begin{eqnarray}
\vec{u}_i^{\textrm{abc}} &=& \frac{V_{dc}+V_O}{6}\left[ \begin{array}{ccc} 
2 & -1 & -1 \\
-1 & 2 & -1 \\
-1 & -1 & 2 
\end{array}\right] \vec{S}_i^{\textrm{abc}} \notag \\
&=& \mathbf{T}_l \vec{S}_i^{\textrm{abc}},
\label{eqn:SU}
\end{eqnarray}
where $V_{dc} = V_{dc1}+V_{dc2}$ is the DC-link voltage and $V_O=V_{dc1}-V_{dc2}$ the DC-link balance.
Their dynamics are given by
\begin{eqnarray}
\dot{V}_{dc} &=& \frac{1}{C}(\vec{S}_m^{\textrm{abc}})^T
 \vec{i}^{\textrm{abc}}_m-\frac{1}{C}(\vec{S}_n^{\textrm{abc}})^T\vec{i}_n^{\textrm{abc}},\\
\dot{V}_O &=&  \frac{1}{C}|\vec{S}_m^{\textrm{abc}}|^T \vec{i}_m^{\textrm{abc}}-\frac{1}{C}|\vec{S}_n^{\textrm{abc}}|^T\vec{i}_n^{\textrm{abc}},
\label{eqn:vos}
\end{eqnarray}
where $\vec{i}_i^{\textrm{abc}}$, $i\in\{m,n\}$ is the Three-phase current flowing through the converter in the system, (see Fig. \ref{fig:sys}).
For the given system the currents have their own dynamics depending on the connecting components which are discussed in the following.
\subsection{Grid side RL-Filter}
On the load side, the converter is connected to a grid with an RL-Filter which can be modeled in $\alpha\beta$-coordinates \cite{PI}.
The relationship between $\alpha\beta$-coordinates and abc-coordinates is given by an inverse Clarke-transform:
\begin{eqnarray}
\vec{i}_i^{\alpha\beta} = \sqrt{\frac{2}{3}}\cdot \left[\begin{array}{ccc}
1 & -\frac{1}{2} & -\frac{1}{2} \\
0 & \frac{\sqrt{3}}{2} & -\frac{\sqrt{3}}{2}
\end{array}\right]\vec{i}_i^{\textrm{abc}} =\mathbf{T}\vec{i}_i^{\textrm{abc}},  \\
\label{eqn:Ttrafo}
\end{eqnarray}
for $i \in \{m,n\}$.
On the grid side the system state is defined as the load side currents $\mathbf{x}_n=\vec{i}^{\alpha\beta}_n$ and the input as voltage $\mathbf{u}_n=\vec{S}^{\textrm{abc}}_n$.
The dynamics of the currents $\vec{i}_n^{\textrm{abc}}$ flowing through the RL-filter are then given by
\begin{eqnarray}
\dot{\vec{i}}^{\alpha\beta}_n  &=& (-\frac{R_n}{L_n})\vec{i}^{\alpha\beta}_n + \frac{1}{L_n} \vec{u}^{\alpha\beta}_n - \frac{1}{L_n} \vec{e}^{\alpha\beta}_n \notag \\
&=&
\mathbf{F}_n \mathbf{x}_n + \mathbf{G}_n\mathbf{u}_n+\mathbf{h}_n.
\label{eqn:gr}
\end{eqnarray}
Furthermore, the active $P$ and reactive power $Q$ that is transferred form the output $\mathbf{y}_n$ of the system:
\begin{eqnarray}
\mathbf{y}_n &=& \left[\begin{array}{c}P \\ Q \end{array}\right] = 
\left[\begin{array}{cc}
e^{\alpha}_n & e^{\beta}_n \\
e^{\beta}_n & -e^{\alpha}_n 
\end{array}\right] \vec{i}^{\alpha\beta}_n \\
&=& \mathbf{C}_n\mathbf{x}_n.
\label{eqn:pr}
\end{eqnarray}  
\subsection{Wind Turbine with PMSG}
The machine side of the converter is connected to a PMSG wind turbine. 
Using the incoming wind, the wind turbine absorbes the power $P_w(t)$ which is given by \cite{Wind,MyThesis,FG3}:
\begin{equation}
P_w(t) = \frac{1}{2}\rho A v_w^3(t) C_p(t).
\label{eqn:wt}
\end{equation}
Here $A$ is the cross section of the turbine, $\rho$ is the wind density,  $v_w$ the wind speed and $C_p$ is the power coefficient of the wind turbine.
In this paper it is assumed that $P_w(t)$ is entirely transformed to mechanical power for the generator $P_g(t)$:
\begin{equation}
P_w(t) = P_g(t) = T_m(t) \omega_m(t),
\label{eqn:wtge}
\end{equation}
where $\omega_m(t)$ is the mechanical generator speed and $T_m(t)$ is the torque.
With the help of the angular momentum we now formulate a relationship between $\omega_m(t)$ and the torque $T_m(t)$:
\begin{equation}
J \dot{\omega}_m (t) = T_m(t)-N_p\psi_{\textrm{pm}}i^{\textrm{q}}_{\textrm{m}}(t)+(\omega_m)(t).
\label{eqn:gete}
\end{equation}
The mechanical generator speed $\omega_m(t)$ is related to the electric speed $\omega_e(t)$ by the number of pole pairs $p$,
\begin{equation}
 \omega_e = p \omega_m.
 \end{equation}
The reference torque $T^{\textrm{ref}}_e$ of the PMSG can be computed as
\begin{equation}
T^{\textrm{ref}}_e = \begin{cases} K^{\textrm{opt}}_1 \omega_m, \quad \textrm{if} v_w \textrm{is unavailale} \\ K^{\textrm{opt}}_2 v_w, \quad \textrm{otherwise}. \end{cases}
\end{equation}
Definitions for $K^{\textrm{opt}}_1$ and  $K^{\textrm{opt}}_2$ can for example be found in \cite{MyThesis}.
For simplicity, the PMSG is modeled in rotating $dq$-coordinates.
The relationship between the stationary $\alpha\beta$-coordinates and rotating $dq$-coordinates is given by a Park-transform:
\begin{eqnarray}
 \vec{i}_i^{dq}= \left[\begin{array}{cc} 
\cos (\theta) & \sin (\theta) \\
- \sin (\theta) & \cos (\theta)
\end{array}\right]\vec{i}_i^{\alpha\beta} = \mathbf{P}\vec{i}_i^{\alpha\beta},
\label{eqn:Ptrafo}
\end{eqnarray}
for $i \in \{m,n\}$.
 The dynamics of the converter in $dq$-coordinates are given by
\begin{eqnarray}
\dot{\vec{i}}^{dq}_m &=& 
\left[\begin{array}{cc} 
-\frac{R_s}{L_s} &  \omega_e \\
- \omega_e & -\frac{R_s}{L_s}
\end{array}\right] \vec{i}^{dq}_m+ \notag \\ 
&\quad& +\left[\begin{array}{cc}\frac{1}{L_s} & 0 \\ 0 & \frac{1}{L_s}\end{array}\right]\vec{u}^{dq}_m -\frac{ \psi_{\textrm{pm}}}{L_s}\left[\begin{array}{c} 0 \\\omega_e\end{array}\right] \notag\\
&=& \mathbf{F}_m \mathbf{x}_m + \mathbf{G}_m \mathbf{u}_m + \mathbf{h}_m,
\label{eqn:ge}
\end{eqnarray}
defining the system state $\mathbf{x}_m = \vec{i}_m^{dq}$ and input $\mathbf{u}_m=\vec{S}_m^{abc}$.
On the machine side, the system output $\mathbf{y}_m$ shall be the state $\mathbf{x}_m$ which means that $\mathbf{C}_m$ is an identity matrix for the remaining paper.
The system model is summarized in equation \eqref{eqn:pr}, \eqref{eqn:ge} and \eqref{eqn:vos}.
To ease the notation, the sampled versions of $\mathbf{x}(t)$ and $\mathbf{x}(t+T_s)$ will be denoted as $\mathbf{x}[k]$ and $\mathbf{x}[k+1]$ for the remainder of the paper, where $T_s$ is the sampling time of the controller.

\begin{figure}[!htbp]
\begin{center}
\subfigure[Block diagram of Predictive Current Control \cite{PCC}. The Three-phase current, machine speed, and DC-link voltage are taken as input.
The references are the reference machine side $d-$ axis current $i^{d,\textrm{ref}}_m = 0$,
	and the reference machine side $q-$ axis current $i^{q,\textrm{ref}}_m=T^{\textrm{ref}}_e/1.5p\psi_{PM}$
.]{ 
			\includegraphics[scale=0.3]{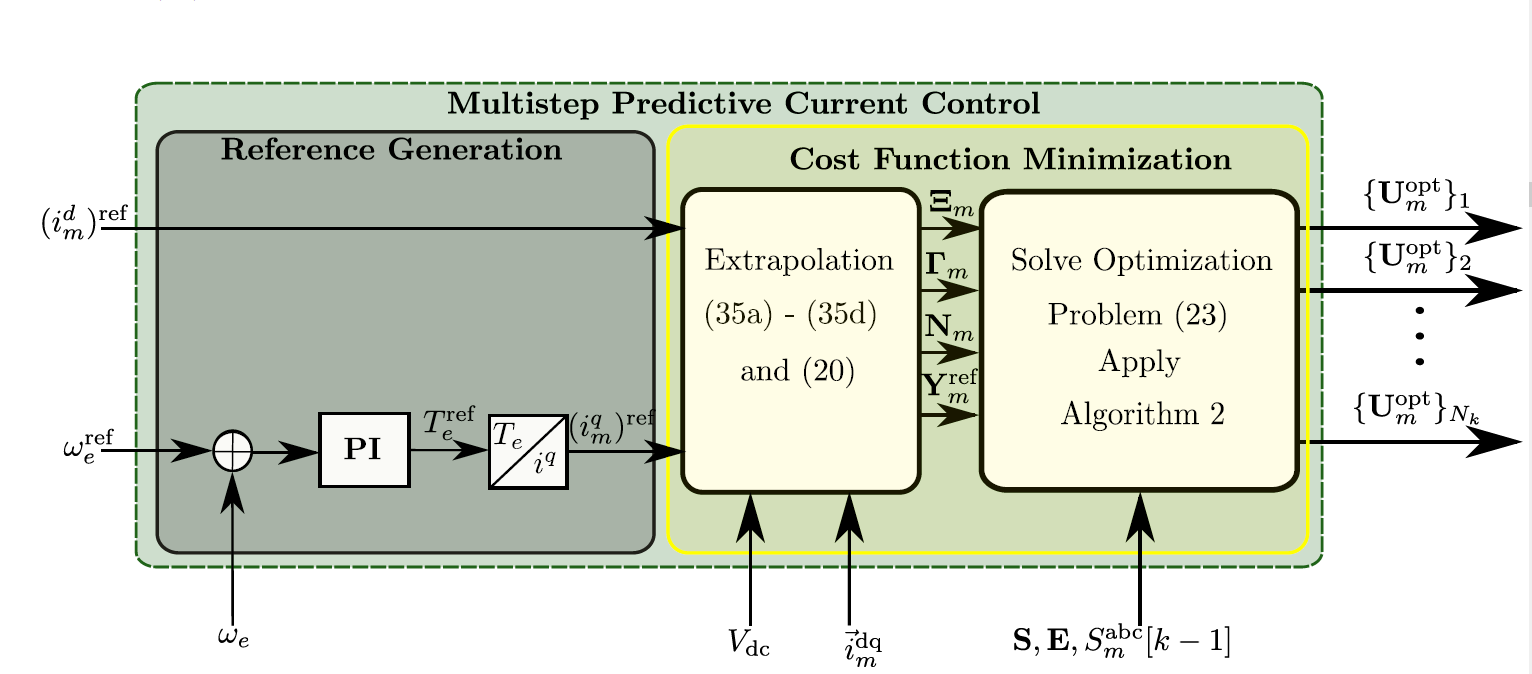}
	\label{fig:PCC}}\\
\subfigure[Block diagram of Predictive Power Control \cite{PPC}. The Three-phase current,  machine speed, machine Torque and DC-link voltage are taken as input. The corresponding reference quantities are the reference grid side reactive power  $Q^{\textrm{ref}} =0$ and
the reference grid side active power $P^{\textrm{ref}} = V_{\textrm{dc}}\cdot I_g^{\textrm{ref}} + \omega^{\textrm{ref}}_m T_e^{\textrm{ref}}$, where $I^{\textrm{ref}}_g$ is estimated using a PI-controller
.]{
			\includegraphics[scale=0.3]{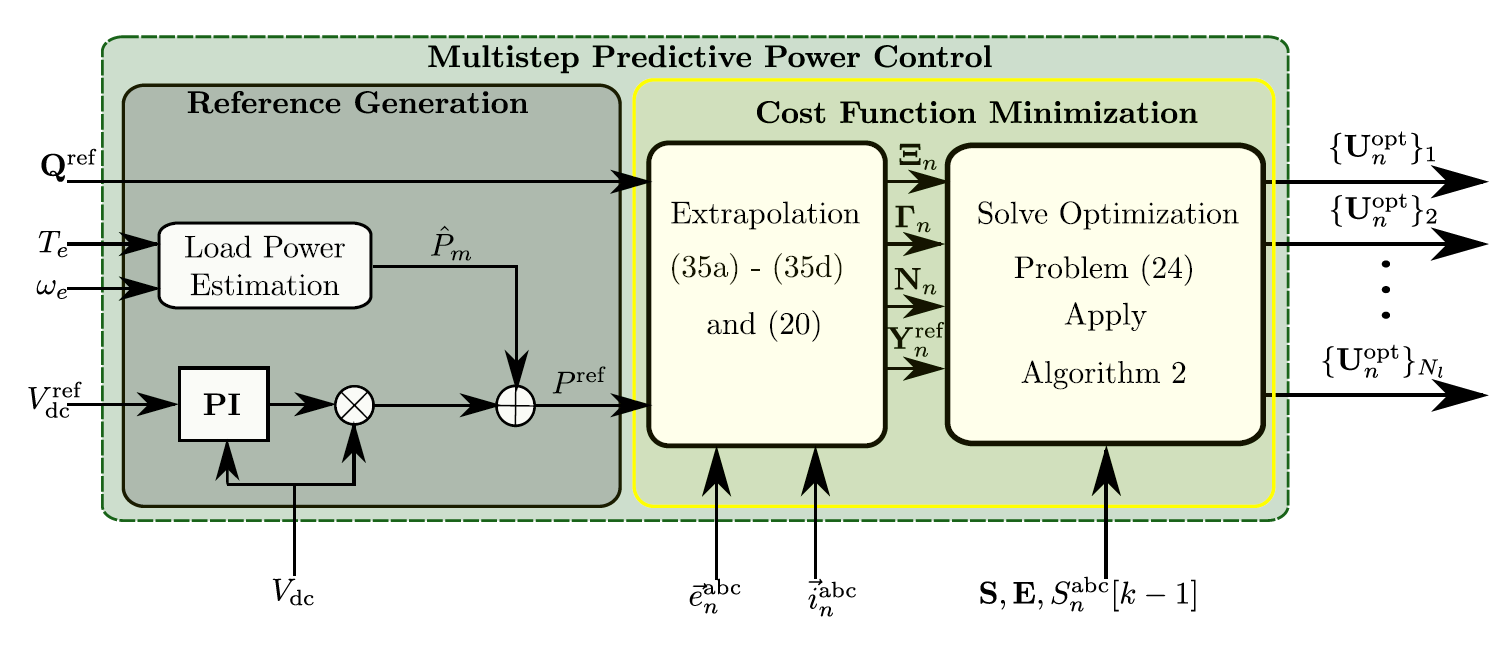}
	\label{fig:PPC}}
	\label{fig:PPP}
	\caption{The basic multistep MPC strategies which are used in this paper: PCC on the machine side and PPC on the grid side. After the best $N_k$ solutions are found, those are evaluated by the DC-link controller.}
	\end{center}
\end{figure}
\vspace{-2ex}
\begin{figure}[!htbp]
    \centering
    \includegraphics[scale=0.4]{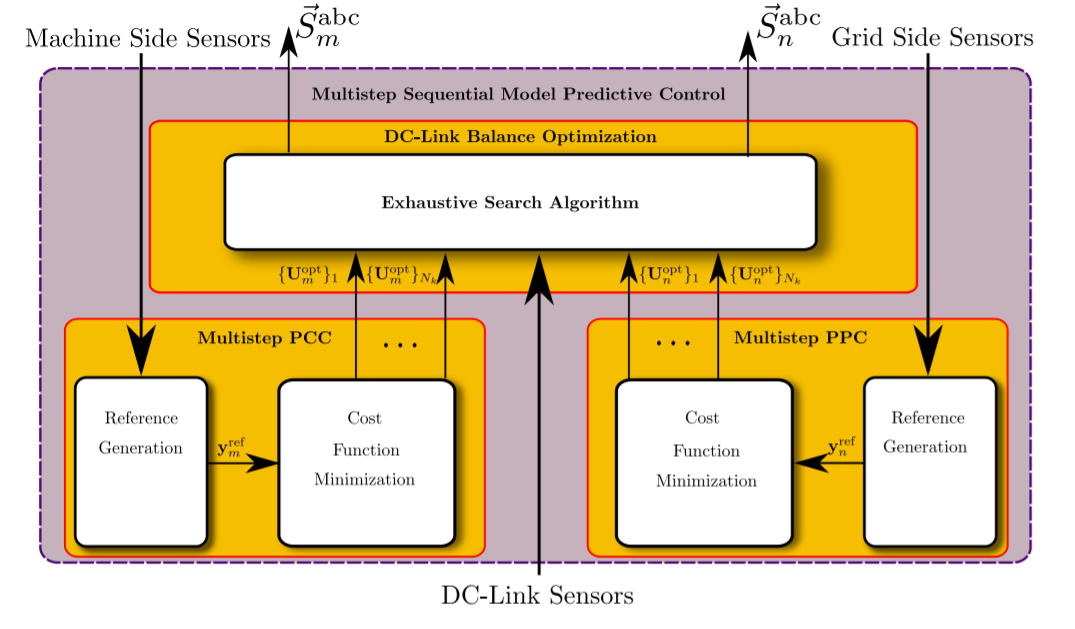}
    \caption{Block diagram of the proposed MPC scheme. After the input signals are recorded the three-phase current trajectory is extrapolated to the next sampling step. The extrapolated three-phase current is then compared to a reference. The optimal input at the next sampling step is selected as the one that yields an output which is closest to the reference.}
    \label{fig:PPV}
\end{figure}
\section{Optimization Problem}\label{sec:problem}
The objective is to control the given system, namely the wind turbine using Predictive Current Control (see Fig. \ref{fig:PCC}), the grid side using Predictive Power Control (see Fig. \ref{fig:PPC}), and the DC-link using DC-link balance control (see Fig. \ref{fig:PPV}).

\subsection{Single Step Prediction}
In this paper, the system outputs are predicted using the Euler approximation
\begin{equation}
\mathbf{x}[k+1] = \mathbf{x}[k]+T_s \dot{\mathbf{x}}[k].
\end{equation}
Since the system equations \eqref{eqn:ge} and \eqref{eqn:gr} are linear the prediction model is formulated as a linear equation of the form
\begin{equation}
\mathbf{x}_i[k+1] = \mathbf{A}_i\mathbf{x}_i[k]+\mathbf{B}_i\mathbf{u}_i[k]+\mathbf{n}_i,
\end{equation}
where
\begin{subequations}
	\begin{eqnarray}
	\mathbf{A}_i &=& \mathbf{I}+T_s\mathbf{F}_i\\
	\mathbf{B}_i &=& T_s\mathbf{G}_i\\
	\mathbf{n}_i &=& T_s\mathbf{h}_i.
	\end{eqnarray}
\end{subequations}
The computation of a prediction model for the DC-link balance 
$V_O = V_{\textrm{dc}_1}-V_{\textrm{dc}_2}$
can be found in a similar way,
\begin{eqnarray}
{V}_{O}[k+1] &=& V_O+\frac{T_s}{C}|\mathbf{u}_m[k]|^T
\mathbf{T}^{\dagger}\mathbf{P}^T\mathbf{x}_m \notag \\
&\quad& -\frac{T_s}{C}|\mathbf{u}_n|^T\mathbf{T}^{\dagger}\mathbf{x}_n,
\label{eqn:VOP}
\end{eqnarray}
where $\mathbf{T}^{\dagger}$ denotes a Moore-Pennrose Pseudo inverse of $\mathbf{T}$.
The prediction model of the DC-link balance is nonlinear with respect to the system input due to the product between the input $\mathbf{u}$ and state $\mathbf{x}$ which again depends on $\mathbf{u}$.
The reference outputs $\mathbf{y}^{\textrm{ref}}_m$ and $\mathbf{y}^{\textrm{ref}}_n$ used in this paper are given in Fig. \ref{fig:PPP}.

\subsection{Multi Step Prediction}
The performance of model predictive control can be improved by predicting the behavior of the system over more than one sampling step \cite{LH}. Let $N_h$ be the number of sampling steps of interest. Then the multistep system state is defined as follows \cite{FG3}:
\begin{equation}
\mathbf{X}_i[k] = \left[\begin{array}{cccc}\mathbf{x}^T_i[k] & \mathbf{x}^T_i[k+1] & ... & \mathbf{x}^T_i[k+N]\end{array}\right]^T,
\end{equation}
 for $i \in \{m,n\}$.
In the same way, $\mathbf{Y}_i$, $\mathbf{U}_i$, $\mathcal{V}_O$ and $\mathbf{Y}^{\textrm{ref}}_i$ are defined.
This yields to the multistep input-output relationship given as
\begin{equation}
\mathbf{Y}_i[k+1]  = \mathbf{\Xi}_i \mathbf{U}_i[k]+\mathbf{\Gamma}_i \mathbf{x}_i[k]+\mathbf{N}_i.
\label{eqn:MmIO}
\end{equation}
The matrices $ \mathbf{\Xi}_i, \mathbf{\Gamma}_i$ and $\mathbf{N}_i$ are given as:
\begin{subequations}
	\fontsize{8.5pt}{8.5pt}{
	\begin{eqnarray}
	\mathbf{\Xi}_i &=& \left[
	\begin{array}{cccc}
	\mathbf{C}_i\mathbf{B}_i & \mathbf{0} & ... & \mathbf{0}\\
		\mathbf{C}_i\mathbf{A}_i \mathbf{B}_i &  \mathbf{B}_i  & ... & \mathbf{0} \\
	\vdots & \vdots & \ddots & \vdots \\
		\mathbf{C}_i\mathbf{A}^{N-1}_i \mathbf{B}_i &  	\mathbf{C}_i\mathbf{A}^{N-2}_i \mathbf{B}_i & ...
	& 	\mathbf{C}_i\mathbf{B}_i 
	\end{array}
	\right] , \\
	\mathbf{\Gamma}_i &=& \left[\begin{array}{cccc} 	\mathbf{C}_i\mathbf{A}^T_i &  	\mathbf{C}_i\left(\mathbf{A}^2_i\right)^T & ... & 	\mathbf{C}_i\left(\mathbf{A}^N_i\right)^T \end{array}\right]^T,\\
	\mathbf{N}_i &=& \left[\begin{array}{ccc} 	\mathbf{C}_i\left(\mathbf{n}_i\right)^T  & ... & 	\mathbf{C}_i\left(\sum_{l=0}^{N-1}\mathbf{A}^{l}_i\mathbf{n}_i\right)^T \end{array}\right]^T.
	\end{eqnarray}}
\end{subequations}
Another quantity of interest is the control effort  $\Delta\mathbf{u}$. The control effort is defined as the difference between the current switching state and the next one:
\begin{equation}
\Delta\mathbf{u} =\mathbf{u}[k]-\mathbf{u}[k-1].
\end{equation}
A low control effort results in a reduced switching frequency of the converter and thus lowers the switching losses of the system \cite{Multi}.
The multistep control effort $\Delta\mathbf{U}$ can be formulated from the input $\mathbf{U}$ as well:
\begin{equation}
\mathbf{\Delta}\mathbf{U}_i = \mathbf{S}\mathbf{U}_i[k]-\mathbf{E}\mathbf{u}_i[k-1],
\label{eqn:MmCF}
\end{equation}
where $\mathbf{S}$ and $\mathbf{E}$ are given in the as
\begin{subequations}
\begin{eqnarray}
	\mathbf{S} &=& \left[
	\begin{array}{cccc}
	\mathbf{I} & \mathbf{0} & ... & \mathbf{0}\\
	-\mathbf{I} &  \mathbf{I}  & ... & \mathbf{0} \\
	\vdots & \vdots & \ddots & \vdots \\
	\mathbf{0} &  \mathbf{0} & ...
	& \mathbf{I} 
	\end{array}
	\right], 
	\mathbf{E} = \left[\begin{array}{c} \mathbf{I} \\  \mathbf{0}\\ \vdots \\ \mathbf{0} \end{array}\right]
	\end{eqnarray}
	\end{subequations}
for $i \in \{m,n\}$.
The formulation of a multistep prediction model of the DC-link balance depending on a general input is difficult (see \cite{FG3}).
If however only a certain input is considered, it is possible to predict the DC-link balance step-by-step over the horizon by successively applying equation \eqref{eqn:VOP}.
\subsection{Formulation of the Optimization Problem}
The goal is to control all outputs of the PMSG wind turbine system. In order to achieve this, three cost functions $J_m, J_n$ and $J_O$ 
\begin{eqnarray}
J_m &=& \|\mathbf{Y}_m-\mathbf{Y}^{\textrm{ref}}_m\|_2^2+\lambda \|\Delta\mathbf{U}_m\|_2^2 \label{eqn:J_m} \\
J_n &=& \|\mathbf{Y}_n-\mathbf{Y}^{\textrm{ref}}_n\|_2^2+\lambda \|\Delta\mathbf{U}_n\|_2^2 \label{eqn:J_n} \\
J_O &=& \|\mathcal{V}_O\|_2^2. \label{eqn:J_o} 
\end{eqnarray}
need to be minimized.
In conventional MPC, the total cost function $J$ of the system is usually defined as a sum of all 3 subcostfunctions, (see e.g. \cite{MyThesis}):
\begin{equation}
    J = J_m+J_n+J_O.
\end{equation}
The control algorithm then aims to find inputs that minimize this cost function and are valid switching states of the converter:
\begin{eqnarray}
    \mathbf{U}_m^{\textrm{opt}},\mathbf{U}_n^{\textrm{opt}} &=& \arg \min (J(\mathbf{U}_m,\mathbf{U}_n)) \notag \\ &\quad& \textrm{s.t. }  \mathbf{U}_m^{\textrm{opt}},\mathbf{U}_n^{\textrm{opt}} \in \{-1,0,1\}^{3N_h}.
    \label{eqn:opnonlinear}
\end{eqnarray}
It is noteworthy that the PMSG as well as the RL-load only depend on the inputs on their respective side $\vec{S}_m^{\textrm{abc}}$ or $\vec{S}_n^{\textrm{abc}}$.
The dynamics of the DC-link however depends on the input of both sides $\vec{S}_m^{\textrm{abc}}$ and $\vec{S}_n^{\textrm{abc}}$. \newline
In order to reduce the computational complexity, a sequential MPC approach is selected in this paper:
\begin{enumerate}
	\item Outer loop: Minimize $J_m$ and $J_n$ (Fig. \ref{fig:PCC}, \ref{fig:PPC}).
	Furthermore, the solution must be a valid switching state of the Three-Level NPC Converter on both sides.
	\item Inner loop: Minimize $J_O$ (Fig. \ref{fig:PPV}).
\end{enumerate}
In the outer loop, the ${N_k}$-best solutions of $J_m$ and the ${N_l}$-best solution of $J_n$ are computed in parallel. Those solutions are then passed on to the minimization of $J_O$.
Thus cost functions \eqref{eqn:J_m}, \eqref{eqn:J_n} and \eqref{eqn:J_o} can be decomposed into 3 sub-optimization problems
\begin{enumerate}
	\item
Using equation \eqref{eqn:MmIO} and \eqref{eqn:MmCF} allows to formulate the outer loop optimization problems which yields the optimal system input on the machine side:
\begin{eqnarray}
\{\mathbf{U}_{m,k}^{\textrm{opt}}\}_{k=1}^{N_k} &=& \arg\min \|\mathbf{Y}_m-\mathbf{Y}^{\textrm{ref}}_m\|+\lambda \|\Delta\mathbf{U}_m\| \notag \\ &\quad& \textrm{s.t.}\mathbf{U}_m\in \{-1,0,1\}^{3N_h}
\label{eqn:OPM}
\end{eqnarray}
\item 
The outer loop optimization problem on the net side is given by:
\begin{eqnarray}
\{\mathbf{U}_{n,l}^{\textrm{opt}}\}_{l=1}^{N_l} &=& \arg\min \|\mathbf{Y}_n-\mathbf{Y}^{\textrm{ref}}_n\|+\lambda \|\Delta\mathbf{U}_n\| \notag \\ &\quad& \textrm{s.t.}\mathbf{U}_n\in \{-1,0,1\}^{3N_h}
\label{eqn:OPN}
\end{eqnarray}
\item The inner loop optimization problem is then given by
\begin{eqnarray}
\mathbf{U}_{n}^{\textrm{opt}},\mathbf{U}_{m}^{\textrm{opt}} &=& \arg\min \|\mathcal{V}_O\|  \\ &\quad& \textrm{s.t.}\mathbf{U}_n\in \{\mathbf{U}_{n,l}^{\textrm{opt}}\}_{l=1}^{N_l}, \notag \\ &\quad& \quad \mathbf{U}_m\in\{\mathbf{U}_{m,k}^{\textrm{opt}}\}_{k=1}^{N_k}. \notag
\label{eqn:OPV}
\end{eqnarray}
\end{enumerate}
The optimization problem is summarized in \eqref{eqn:OPM}, \eqref{eqn:OPN} and \eqref{eqn:OPV}. 
All three problems are least-squares mixed-integer optimization problems.
Since \eqref{eqn:OPM} minimizes only $J_m$, \eqref{eqn:OPN} minimizes only $J_n$, the optimal solution for \eqref{eqn:opnonlinear} can be lost during this optimization stage.
In this case, the overall algorithm only provides a locally optimal solution that lies within the solutions of the first optimization stage.
Using $\lambda$, a trade-off between control performance and control effort can be tuned. 
The selection of $\lambda$ is not trivial and normally achieved based on experience \cite{GenMPC5}. 
Alternatively, a data-driven method for tuning the parameters of MPC schemes \cite{NN} can be used.

\section{Solution of the optimization problem}\label{sec:solution} 
The first two optimization problems \eqref{eqn:OPM}, \eqref{eqn:OPN} are solved with a modified sphere decoding algorithm.
In order to solve them, they have to be modified to a triangular structure.
Therefore we define similarly to \cite{Multi}:
\begin{eqnarray}
\mathbf{Q}_i &=& \mathbf{\Xi_i}^T\mathbf{\Xi_i}+ \lambda \mathbf{S}^{T}\mathbf{S} \\
\mathbf{\Theta}_i &=& \left(\left(\mathbf{\Gamma}_i\mathbf{x}_i[k]+\mathbf{N}_i-\mathbf{Y}_i^{\textrm{ref}}\right)^T\mathbf{\Xi}_i\right)^T \notag \\
&\quad&  - \lambda\left(\left(\mathbf{E}\mathbf{u}[k-1]\right)^T\mathbf{S}\right)^T.
\label{eqn:QT}
\end{eqnarray}
The triangular structure of the system is now achieved by a Cholesky-decomposition of $\mathbf{Q}_i$:
\begin{equation}
\mathbf{Q}_i = \mathbf{H}_i^T\mathbf{H}_i.
\end{equation}
This allows to define the unconstrained solution of the optimization problems 
$\mathbf{U}_{\textrm{unc},i}$:
\begin{equation}
\mathbf{U}_{\textrm{unc},i} = (\mathbf{Q}_i)^{-1}\mathbf{\Theta}_i.
\end{equation}
Furthermore we define
\begin{equation}
\mathbf{\check{U}}_{\textrm{unc},i} = \mathbf{H}_i \mathbf{U}_{\textrm{unc},i}
\end{equation}
for $i \in \{m,n\}$. 
The $N_k$ and $N_l$ best solutions to the optimization problem \eqref{eqn:OPM}, \eqref{eqn:OPN} are now found with the help of a modified sphere decoding algorithm.
This algorithm computes the required solutions in a sequential manner, starting from the best one and moving to the worst one. 
After a solution is obtained, it is removed from the candidate tree. 
\begin{figure}[!htbp]
    \centering
    \subfigure[First Solution]{
    \includegraphics[scale = 0.35]{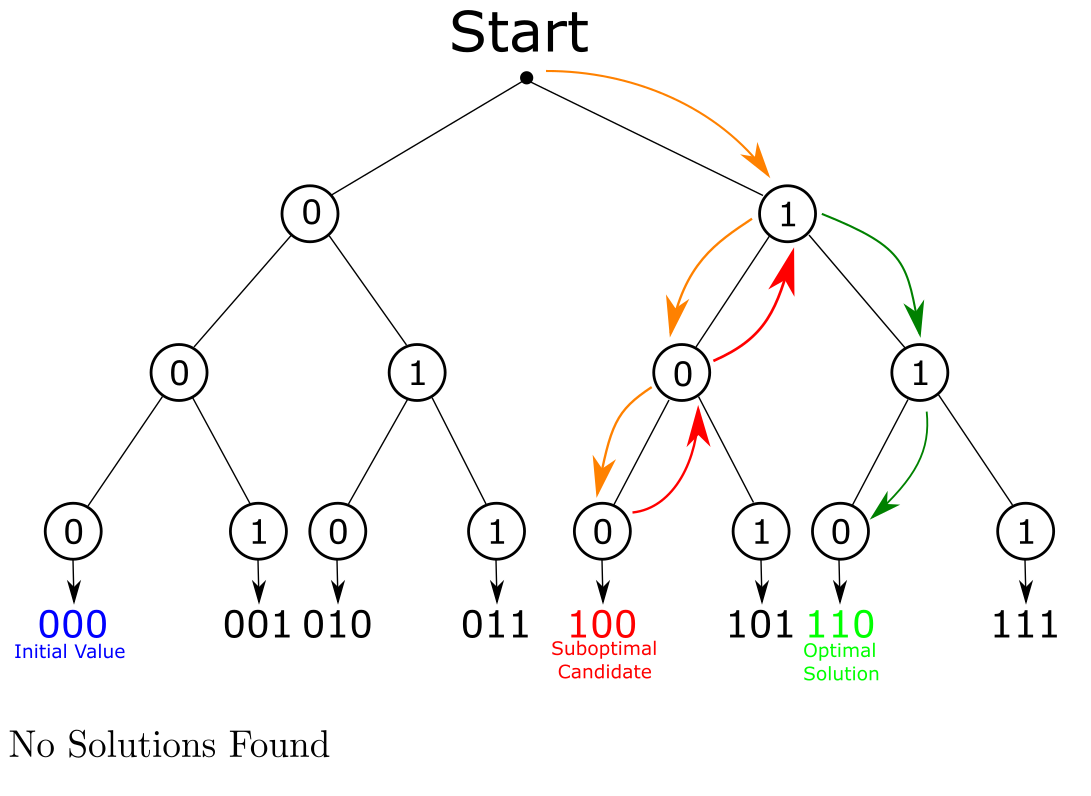}
    }
        \subfigure[Second Solution]{
    \includegraphics[scale = 0.28]{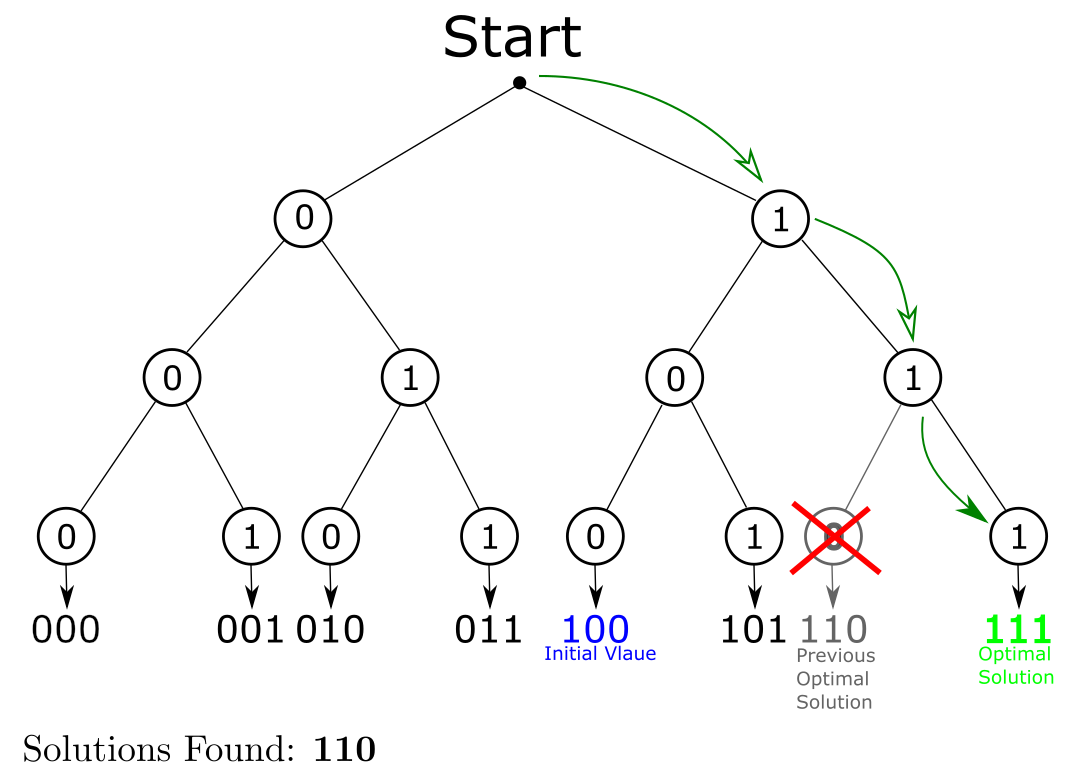}
    }
        \subfigure[Third Solution]{
    \includegraphics[scale = 0.28]{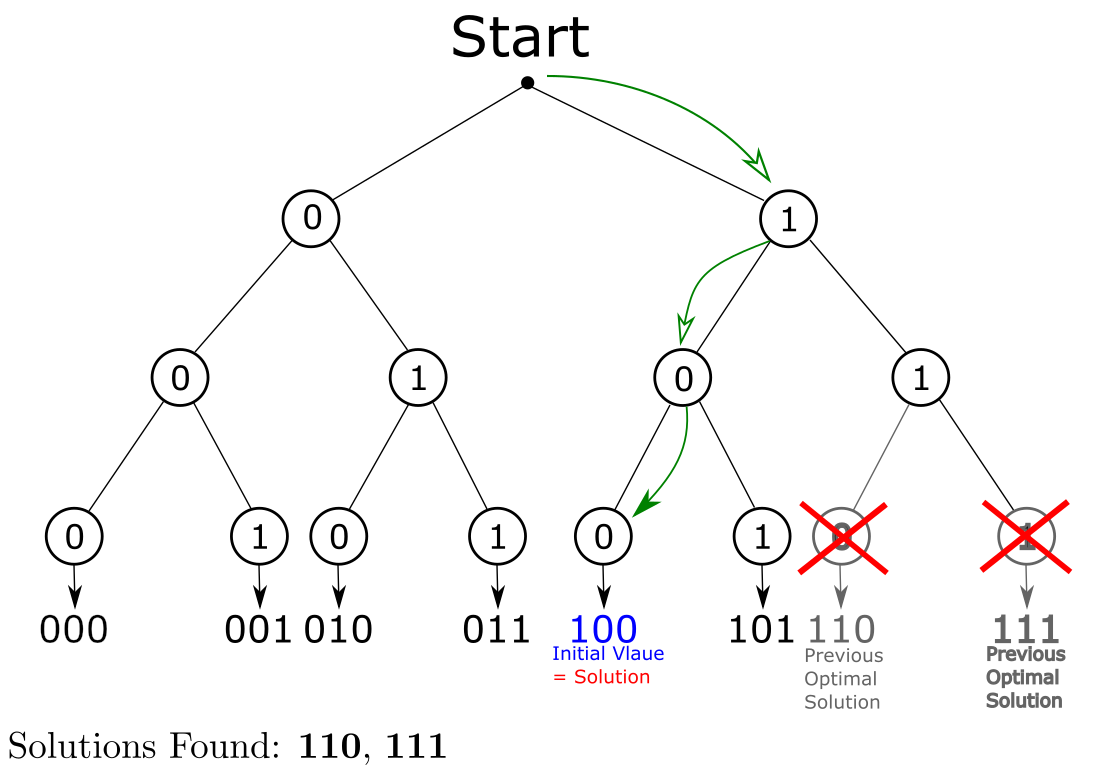}
    }
    \caption{The proposed algorithm for $N_k=3$: Starting from an initial value (blue), the algorithm looks for the best available solution similar to \cite{Multi}. If an optimal solution (green) is found, the second-best solution (red) is saved as well and the algorithm continues. To find the next best solution, the previously optimal candidate is removed from the tree (grey), and the previously optimal candidate is set to the initial value (blue). This procedure is repeated until all $N_k$ solutions are found. }
    \label{fig:KSDExplanation}
\end{figure}
The current second best solution is used as an initial value and the algorithm is called again.
A graphic depiction of the algorithm is shown in Fig. \ref{fig:KSDExplanation}.
Therefore the Algorithm \ref{alg:SD} is called $N_k$ (or $N_l$) times and after each call the previously found solution is removed from the candidate set $\mathcal{U}$:
\begin{algorithm}[h]
	\caption{Proposed sphere decoding algorithm.}\label{alg:SD}
	\begin{algorithmic}[1]
		\Function{$\mathbf{U}^{\textrm{opt}}_1, \mathbf{U}^{\textrm{opt}}_2= $ SD}{$\tilde{\mathbf{U}},\mathbf{H},d^2,k,\rho^2,\mathbf{\check{U}}_{\textrm{unc}},\mathcal{U}$}
		\For{$\mathbf{u}\in \mathcal{U}(k)$}
		\State{$\tilde{\mathbf{U}}_k = \mathbf{u}$}
		\State{$\tilde{d}^2 = \|\mathbf{\check{U}}_{\textrm{unc}}-\mathbf{H}_{k,1:k}\tilde{\mathbf{U}}_{1:k}\|$}
		\If{$\tilde{d}^2 \le \rho^2$} 
		\If{$k > 1$} 
		\State{$\mathbf{U}^{\textrm{opt}}_1,\mathbf{U}^{\textrm{opt}}_2=$ SD(\small{$\tilde{\mathbf{U}},\mathbf{H},\tilde{d}^2,k+1,\rho^2,\mathbf{\check{U}}_{\textrm{unc}},\mathcal{U}$})}
		\Else
		\State{$\mathbf{U}^{\textrm{opt}}_2 = \mathbf{U}^{\textrm{opt}}_1$}
		\State{$\mathbf{U}^{\textrm{opt}}_1 = \tilde{\mathbf{U}}$}
		\State{$\rho^2 = \tilde{d}^2$}
		\EndIf
		\EndIf 
		\EndFor\label{euclidendwhile}
		\State \textbf{return} $\mathbf{U}^{\textrm{opt}}_1$, $\mathbf{U}^{\textrm{opt}}_2$
		\EndFunction
	\end{algorithmic}
\end{algorithm}
The algorithm is initialized by setting $\rho_t^2 = \mathbf{0}$, and $\mathcal{U} = \{-1,0,1\}^{3N}$ for $i \in \{m,n\}$ and $j \in \{1,2,...,N_k\}$.
This algorithm has two differences compared to the one proposed in \cite{Multi}:
\begin{itemize}
	\item It also provides the second best solution $\mathbf{U}^{\textrm{opt}}_2$ which is used as initial value in the next iteration. 
	\item The input set $\mathcal{U}$ shrinks after each iteration.
\end{itemize}
The sphere decoder given in algorithm \ref{alg:SD} is able to find an optimal solution to \eqref{eqn:OPM} or \eqref{eqn:OPN} respectively \cite{Multi}.
Removing this solution from the candidates and calling the algorithm again will this yield the second-best solution.
In this way, the $N_k$ or $N_h$ best solutions can be obtained by repeating this procedure $N_k$ or $N_h$ times.
The whole procedure is summarized in Algorithm \ref{alg:KSD}.
\begin{algorithm}[h]
	\caption{Proposed algorithm for $N_k$-solutions.}\label{alg:KSD}
	\begin{algorithmic}[1]
		\Function{$\{\mathbf{U}^{\textrm{opt}}\}_{i=1}^{N_i}= $ NKSD}{$\tilde{\mathbf{U}},\mathbf{H},\mathbf{\check{U}},\mathcal{U},N_k$}
		\State{Initialize $\tilde{\mathbf{U}},\mathbf{H},d^2,\rho^2,\mathbf{\check{U}}_{\textrm{unc}},\mathcal{U}$}
		\For{$n_i = 1, n_i < N_i, n_i++$}
		\State{$\{\mathbf{U}^{\textrm{opt}}\}_{n_i}, \mathbf{U}^{\textrm{opt}}_2= $ SD($\tilde{\mathbf{U}},\mathbf{H},d^2,1,\rho^2,\mathbf{\check{U}}_{\textrm{unc}},\mathcal{U}$)}
		\State{$\tilde{\mathbf{U}} =\mathbf{U}^{\textrm{opt}}_2$, $\rho=J_i(\mathbf{U}^{\textrm{opt}}_2)$}
		\State{Exclude $\{\mathbf{U}^{\textrm{opt}}\}_{n_i}$ from $\mathcal{U}$}
		\EndFor
		\State \textbf{return} $\{\mathbf{U}^{\textrm{opt}}\}_{i=1}^{N_i}$
		\EndFunction
	\end{algorithmic}
\end{algorithm} 
After that the best solutions $\{\mathbf{U}_{m,k}^{\textrm{opt}}\}_{k=1}^K$ and $\{\mathbf{U}_{n,l}^{\textrm{opt}}\}_{l=1}^L$ are selected and used to solve the third optimization problem. 
Since the third optimization problem is nonlinear, a sphere decoding algorithm cannot be applied here.
For this reason an exhaustive search has to be performed, testing all possible combinations of $\{\mathbf{U}_{m,k}^{\textrm{opt}}\}_{k=1}^{N_k}$ and $\{\mathbf{U}_{n,l}^{\textrm{opt}}\}_{l=1}^{N_l} $. 
In this case, it is required to perform $N_k \cdot N_l$ computations to achieve the best solution. The last step is to select the first 3 entries of $\mathbf{U}_m^{\textrm{opt}}$ and $\mathbf{U}_n^{\textrm{opt}}$ and apply them on their side of the converter respectively. This procedure is repeated after each sampling step.
\section{Results}\label{sec:result}

Table \ref{tab:mac} shows the parameters used for the simulation.
\begin{table}[!htbp]
	\centering
	\caption{Machine, generator and DC-link parameters}
	\label{tab:mac}
	\begin{tabular}{lll}
		\hline
		\hline 
		\textbf{Parameter}                    & \textbf{Symbol}                     & \textbf{Simulation value}                      \\ \hline
		Generator/Net side resistance & $R_m$, $R_n$               & \num{0.1379}\si{\ohm} , \num{0.156} \si{\ohm} \\
		Generator/Net side inductance & $L_m$, $L_n$               & \num{0.019} \si{\henry}, \num{0.020} \si{\henry}       \\
		Rotor permanent magnet flux   & $\psi_{pm}$                & \num{0.42675} \si{\weber}               \\
		DC-link voltage               & $V_{dc}$                   & \num{700} \si{\volt}                   \\
		DC-link capacitance           & $C$                        & \num{1100} \si{\micro \farad}               \\
		Grid side voltage peak        & $\vec{e}_n^{\textrm{abc}}$ & \num {250} \si{\volt}                    \\
		Grid side voltage frequency   & $\omega_n$                 & \num{100}$\pi$ \si{Hz}               \\
		Generator pole pairs          & $p$                      & $3$                        \\
		Weighting factors             & $\lambda$, $\lambda_V$     & $0.1$, $0.02$              \\ 	\hline \hline
	\end{tabular}
\end{table}
The controller sampling period was set to $T_s = 50 \mu$ s for all scenarios.
First, we examine the transient performance for horizon length $N_h=3$, and $N_k=N_l=4$.

\begin{figure}[!htbp]
	\begin{center}
						\subfigure[ $N_h=1, N_k=4$.]{ 
			\includegraphics[angle=0,scale=0.855]{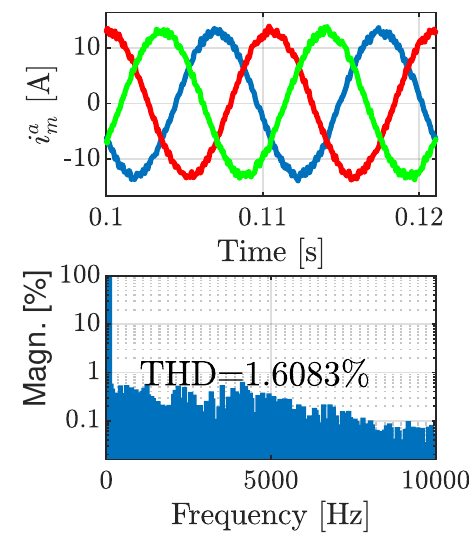}
			\label{Fig7a}
		}
				\subfigure[$N_h=3, N_k=4$.]{ 
		\includegraphics[angle=0,scale=0.855]{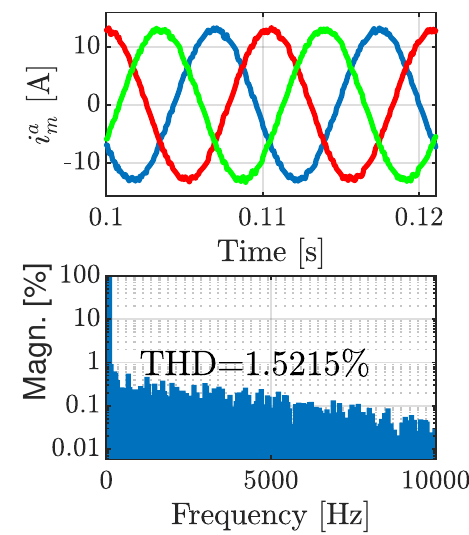}
		\label{Fig7b}
	}
	\end{center}
	\vspace{-2ex}
	\caption{Comparison of different motor currents during the steady-state for different horizon lengths. The top Figures show the Three-phase machine side currents in the time domain. The bottom insets illustrate the frequency domain representation for the Three-phase machine side currents.}
	\label{Steady1}
\end{figure}
The steady-state simulation results further show that the total harmonic distortion (THD) of the current improves with extending the prediction horizon $N_h$ (see Fig. \ref{Steady1}).
While the difference in time domain seems to be small, a decrease of total harmonic distortion can be found for the long-horizon scenario (see Fig. \ref{Fig7a} and Fig. \ref{Fig7b}).
This can be explained since predicting the system behavior over more sampling steps allows identifying switching states that yield good reference tracking over a longer period of time and is confirmed in the literature (see e.g. \cite{Multi2}).
\begin{figure}[H]
\begin{center}
			\includegraphics[scale=0.75]{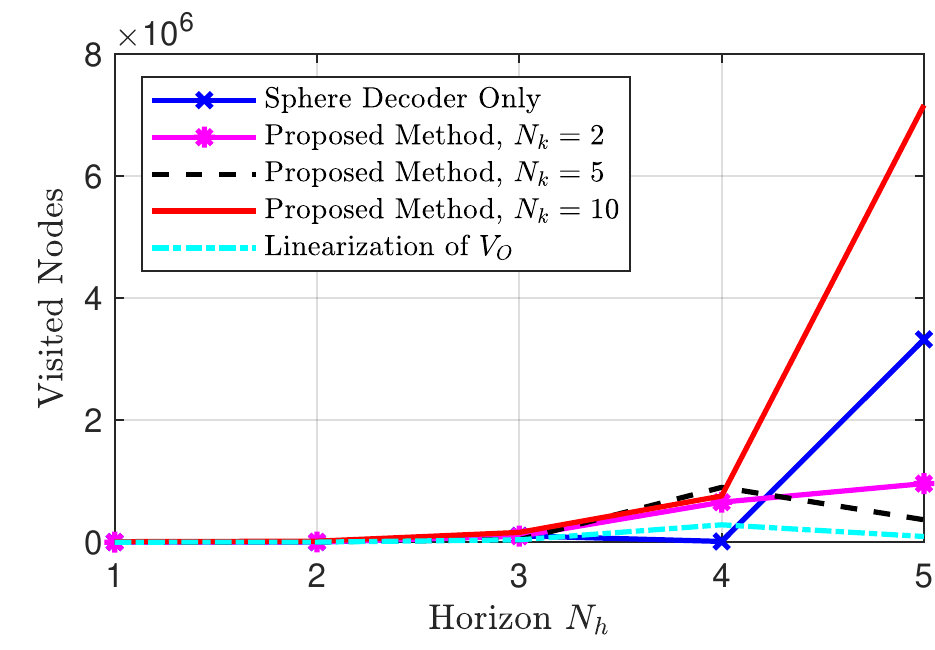}
			\end{center}
			\vspace{-1em}
	\caption{Evaluation of the proposed algorithm in steady-state conditions for different parameter configurations for $N_k$ and $N_h$. For different configurations, the number of nodes that are visited in the search tree is compared. Therefore, the number of nodes is recorded during each sampling step of the simulation. If the algorithm is separated for the machine-and grid side, the number of nodes visited for both sides is added. Shown is the average of the number of nodes visited for the whole simulation duration.
	}
	\label{SSt}
\end{figure}
\begin{figure}[!htbp]
	\begin{center}
			\subfigure[]{ 
	\includegraphics[angle=0,scale=1]{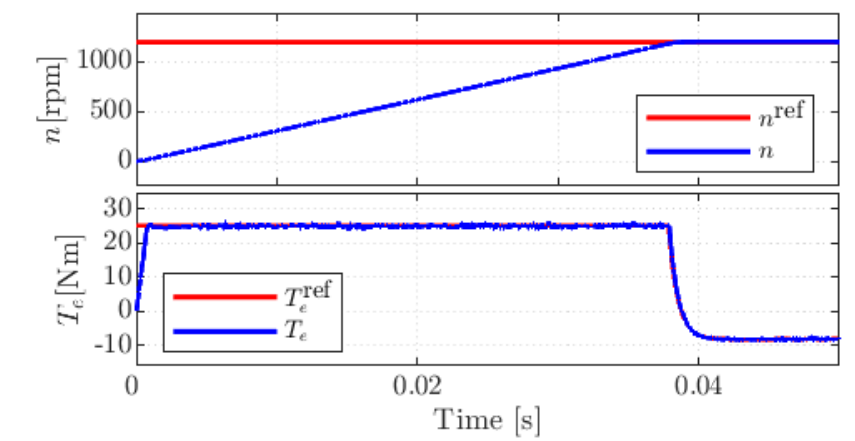}
	\label{Fig8a}
}
\subfigure[]{ 
	\includegraphics[angle=0,scale=1]{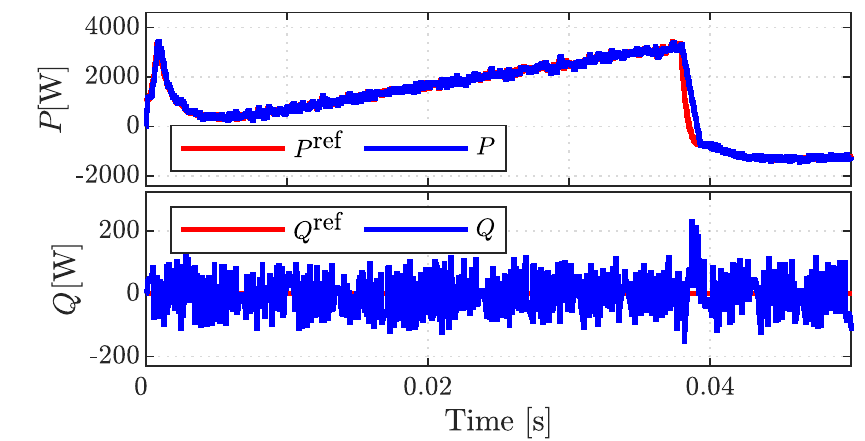}
	\label{Fig8b}
}
		\subfigure[]{ 
	\includegraphics[angle=0,scale=1]{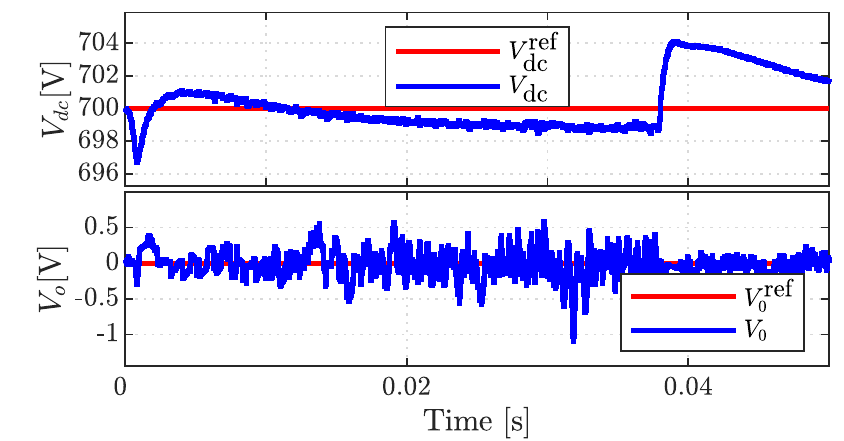}
	\label{Fig8c}
}
	\end{center}
	\caption{Transient simulation results for $N_h=3, N_k=4$. Starting from an idle state, a reference of $1125$rpm is set in the beginning. The results for the machine side quantities are shown in a), the top Figure represents the generator speed and the bottom inset illustrates the generator torque $T_e$. In b) the grid side parameters are shown: The top inset illustrates the active power $P$ and the bottom Figure represents the reactive power $Q$. The DC-link transients are shown in c) where the top inset represents the DC-link voltage $V_{\textrm{dc}}$ and the bottom Figure illustrates the DC-link balance $V_O$.}
	\label{Transients}
\end{figure}
In Fig. \ref{Transients} the simulation results for a transient scenario are shown. 
Fig. \ref{Fig8a} shows that when a reference speed is set, the torque follows its reference closely at the resulting step. The same can be observed for the active power while the performance of the reactive power does not decrease (see Fig. \ref{Fig8b}).  All system variables achieve good reference tracking overall.
Moreover, during the whole simulation, the DC-link balance stays within an interval between $1$V and $-1$V (see Fig. \ref{Fig8c}). 
The DC-link voltage, which is not being controlled directly, is converging to its reference as well with a settling time of roughly $0.03$s.
Since it is not included in the optimization problem, $V_{\textrm{dc}}$ takes a longer time to reach its reference.
This shows that the proposed method is capable of controlling the system and closely following the reference quantities in case of changing conditions.
The results in Fig. \ref{Transients} and Fig. \ref{Steady1} show that the algorithm is capable of meeting the control goal and that existing observations such as improvement of the performance with prediction horizon are valid as well.
\begin{figure*}[!htbp]
	\begin{center}
			\subfigure[]{ 
	\includegraphics[angle=270,scale=0.17]{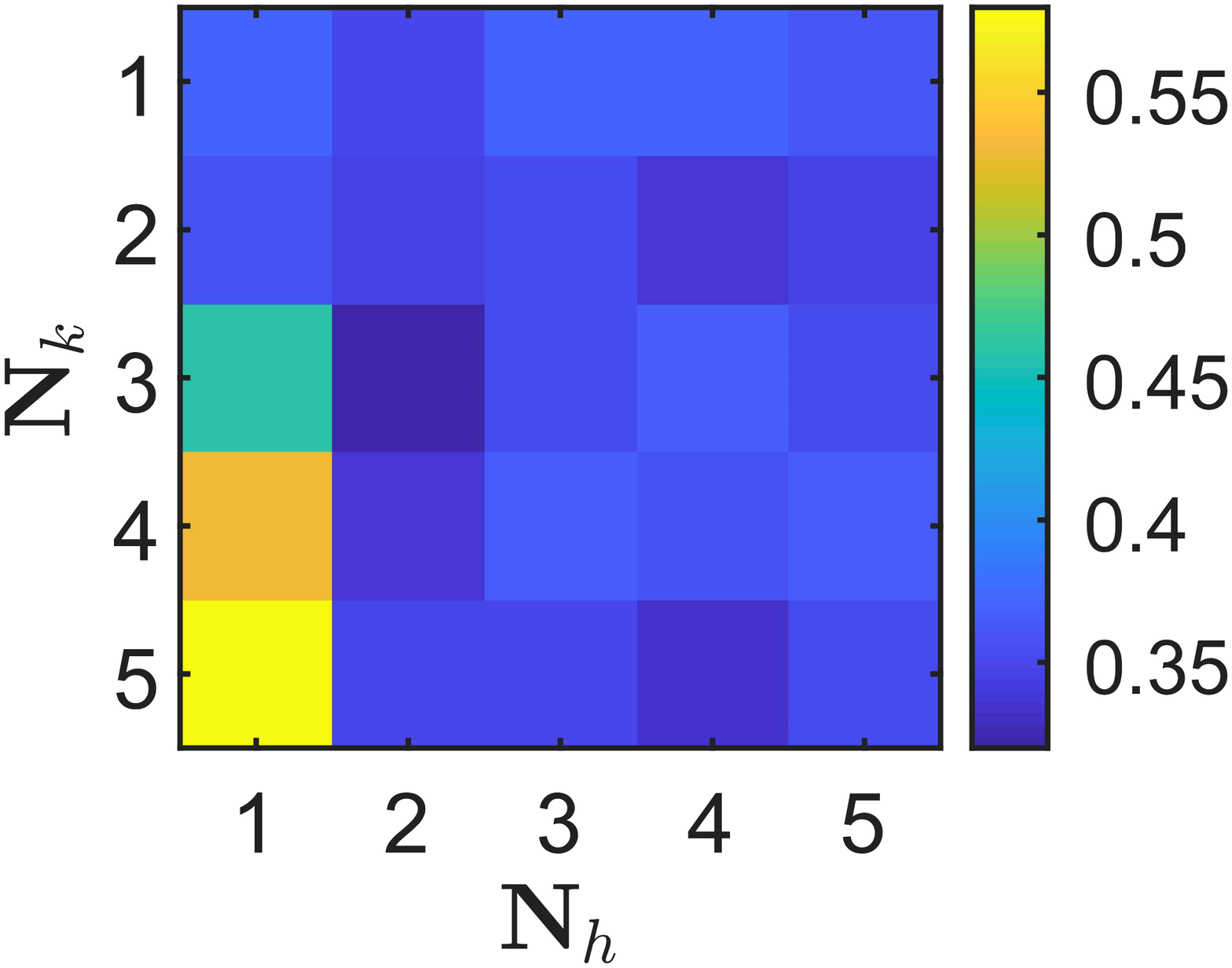}
	\label{Fig9a}
}\hspace{-3.5em}
			\subfigure[]{ 
	\includegraphics[angle=270,scale=0.17]{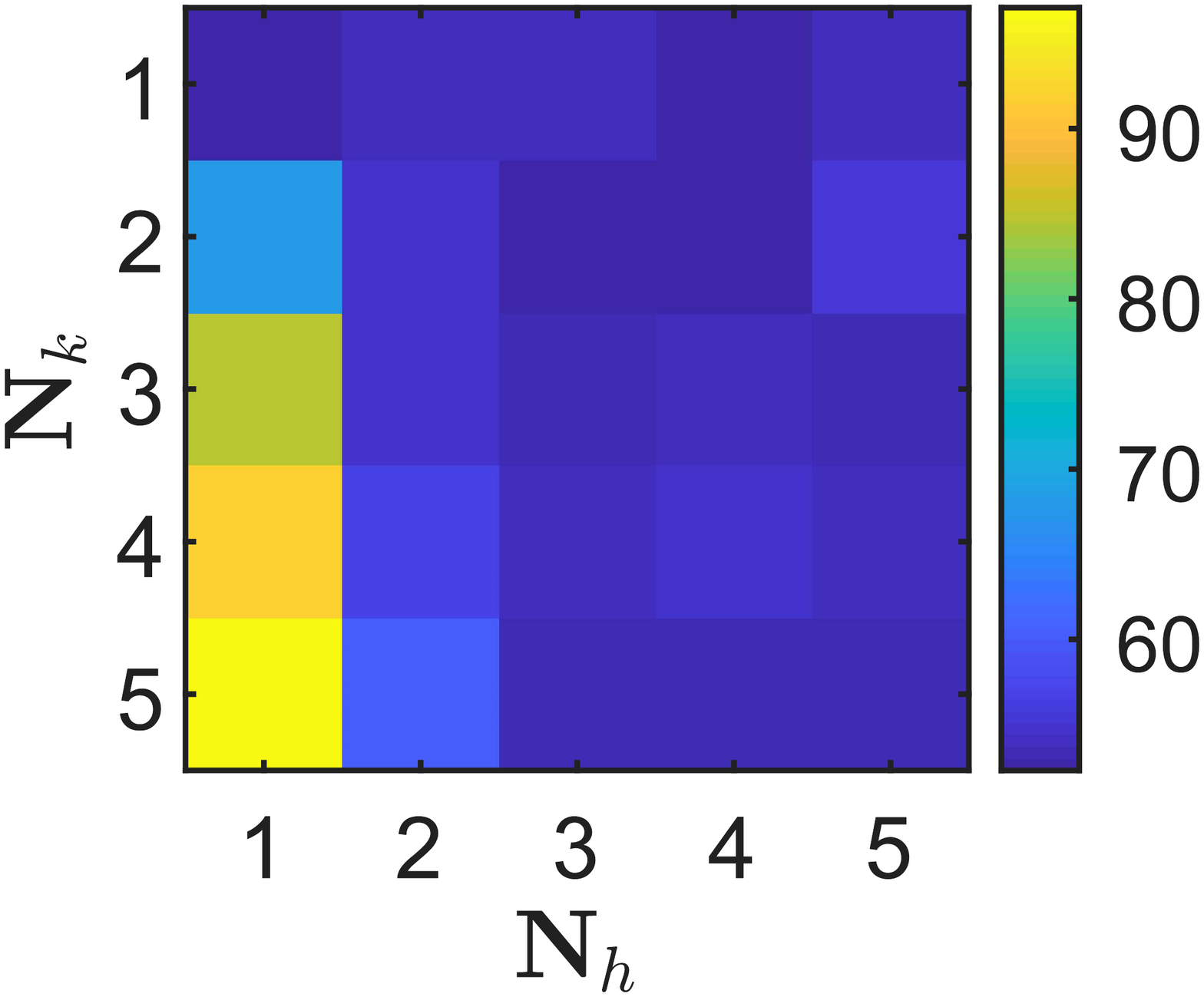}
	\label{Fig9b}
}\hspace{-3.5em}
			\subfigure[]{ 
\includegraphics[angle=270,scale=0.17]{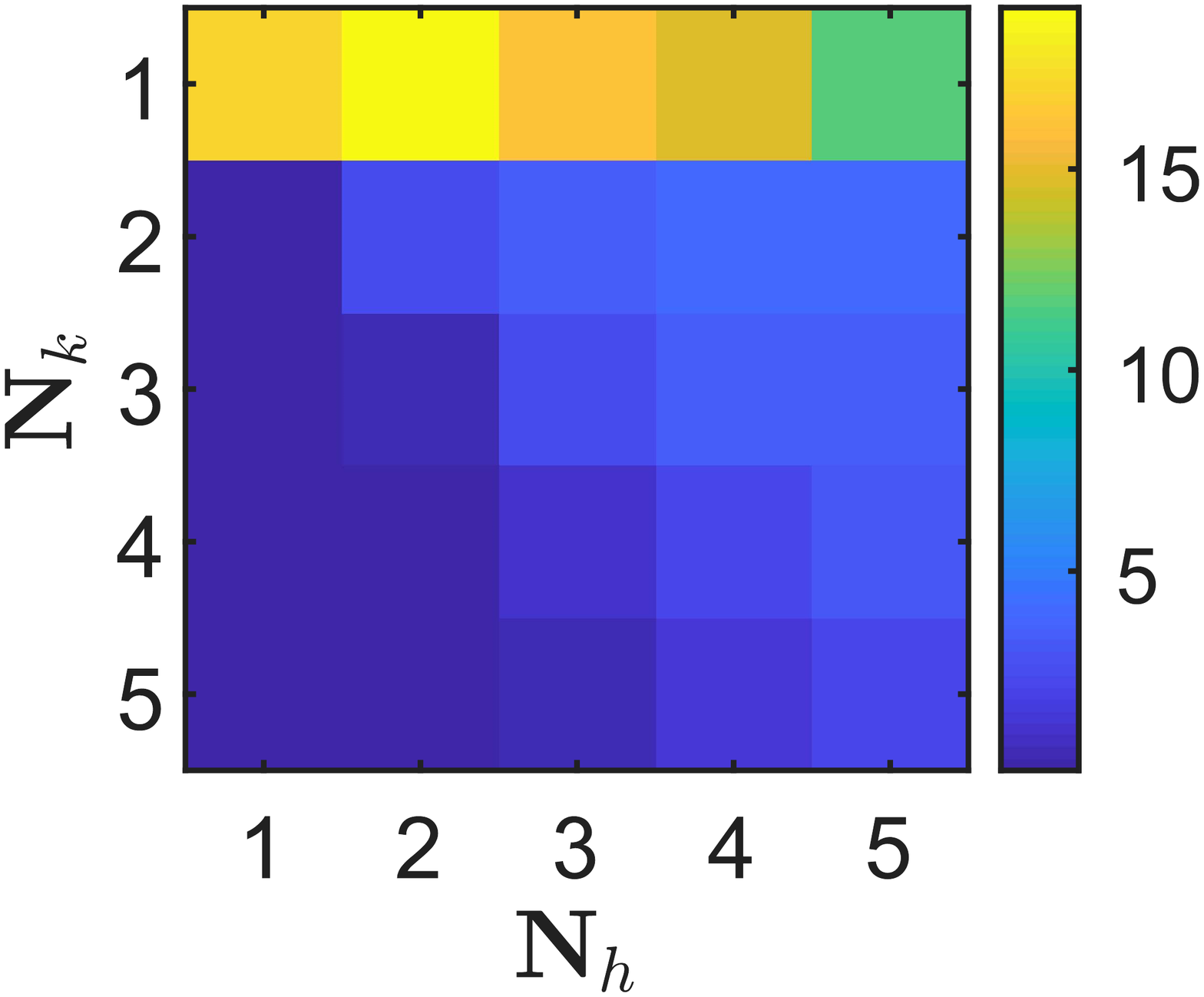}
\label{Fig9c}
}\hspace{-3.5em}
			\subfigure[]{ 
				\vspace{-1ex}
	\includegraphics[angle=270,scale=0.17]{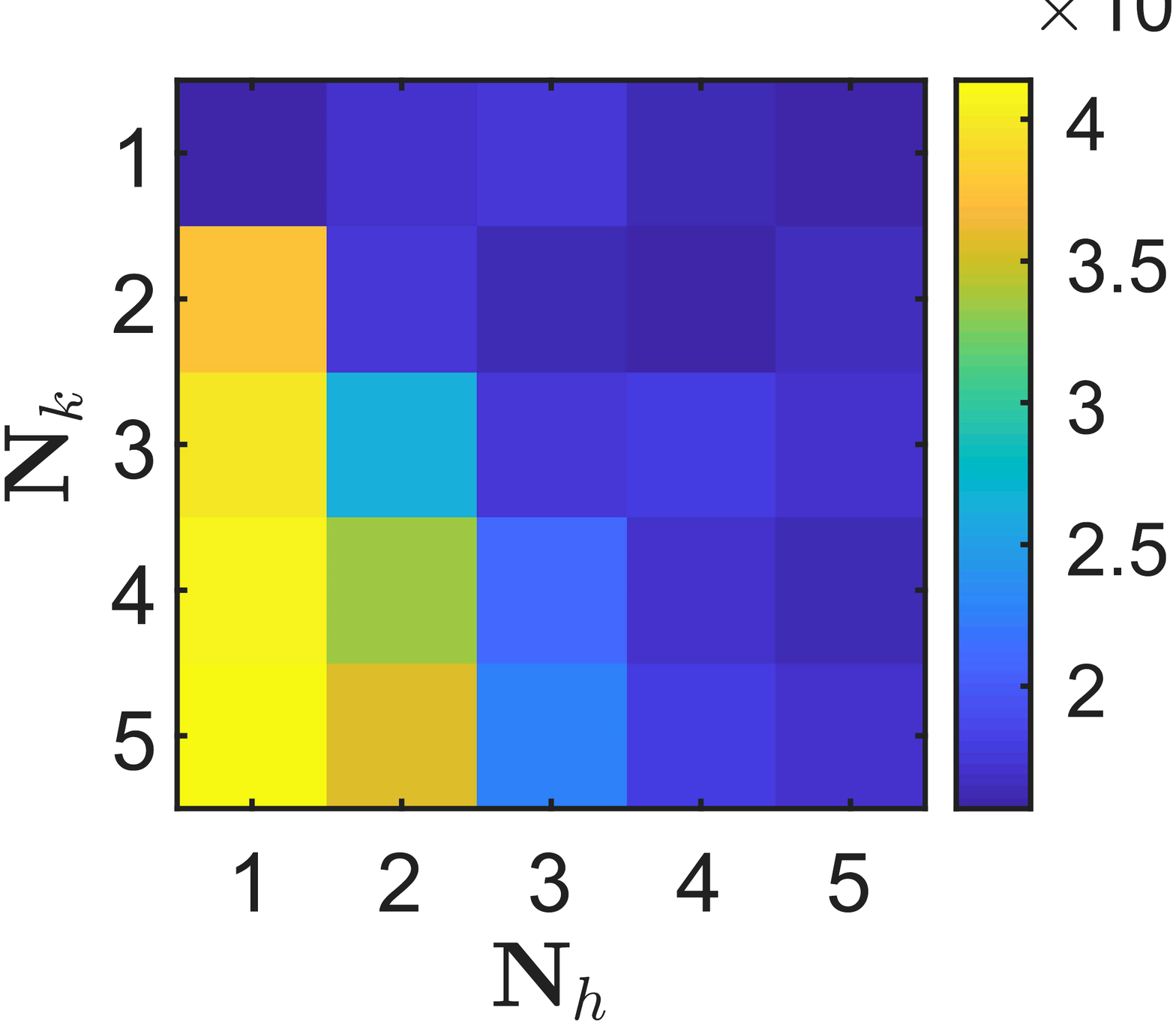}
	\label{Fig9d}
}
\end{center}
		\caption{Steady-state simulation results under varying algorithm parameters $N_h$ and $N_k$. The RMSE of the generator torque $T_e$ is shown in a). In b) the RMSE of the reactive power $Q$ and in c) the RMSE of the DC-link balance $V_{\textrm{O}}$ are shown. Subfigure d) shows the machine side switching frequency in $10$kHz.}
	\label{SSCC}
\end{figure*}
\subsection{Algorithm Parameter Discussion}

The performance of the algorithm depending on the parameters $N_k$ and $N_h$ is given in Fig. \ref{SSCC} and Fig. \ref{SSt}.

Low errors for torque can be achieved by choosing $N_h>3$ as seen in Fig. \ref{Fig9a}. 
Otherwise, the available pool of switching states contains too many values that corrupt the control performance, especially for larger values of $N_k$. For the grid side reactive power, this effect is more extreme, as seen in Fig. \ref{Fig9b}.
This can be explained by the fact that all of those parameters lie in the first control loop, thus they yield similar performance changes with $N_k$ and $N_h$.
On the other hand side, the opposite can be observed for the DC-link balance which is shown in Fig. \ref{Fig9c}.
In this case, small values of $N_k$ lower the performance. With increasing $N_h$ it is possible that all $N_k$ sequences share the same prefix. For this reason, an increase in $N_h$ can be counterproductive for this method if $N_k$ is not large enough.
Since the switching frequency is minimized with the control effort, results for the switching frequency are similar compared to the machine side and grid side (see Fig. \ref{Fig9d}).
Furthermore, it is visible that for the short horizon, the DC-link balance improves even more compared to the long horizon for $N_k=4$. This can be explained since for $N_h=1$, $N_k=4$ the amount of solutions available for DC-link balancing is $4/27\approx 15\%$. On the other hand side, for the long horizon case, the fraction of solutions available for DC-link balancing is $4/27^3 \approx 0.02 \%$.
This shows that the fraction of solutions required to achieve DC-link balancing is very low for the long horizon case.
The results from Fig. \ref{SSCC} can furthermore be used as tuning guidelines of the algorithm if a certain trade-off between control variables should be achieved. 
Finally, the complexity of the sphere decoding part is visualized in Fig. \ref{SSt}. 
It is visible that an increase in $N_k$ has a much smaller impact compared to an increase in horizon length $N_h$, suggesting that larger values of $N_k$ shall be preferred over larger values of $N_h$. 	While for shorter horizons up to $N_h$ the increase in computational cost seems to be small, an exponential increase seems to be visible for reaching $N_h =4 $ and $N_h = 5$.
This indicates that while increasing the $N_h$ is costly in terms of computational complexity, an increase in $N_k$ provides a smaller computational burden.
Thus it is numerically more beneficial to increase $N_k$ to improve the performance over an increase in $N_h$.

\subsection{Comparative Analysis}
\begin{figure*}[!htbp]
	\begin{center}
			\subfigure[]{ 
	\includegraphics[scale=0.55]{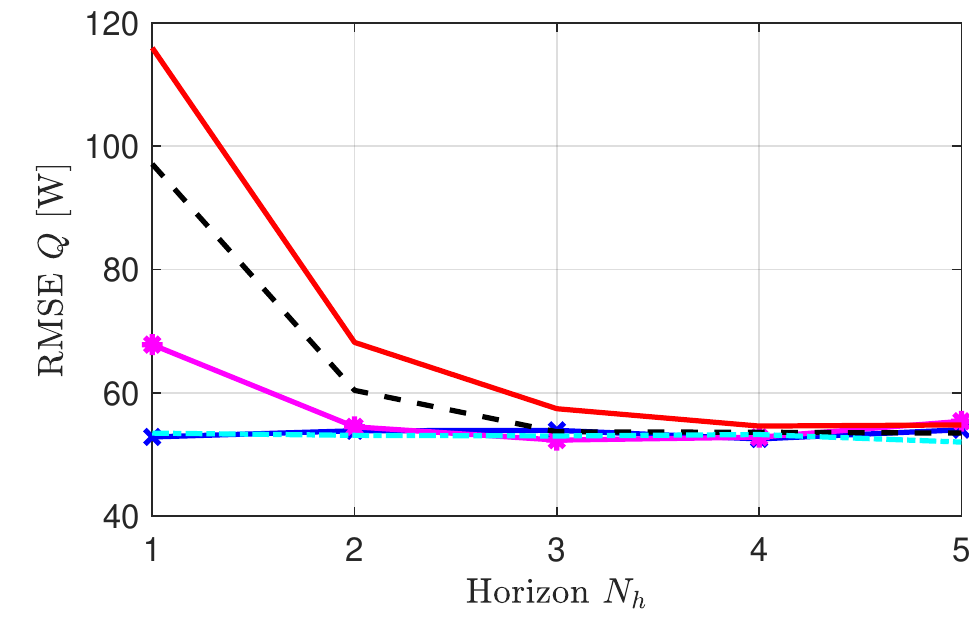}
	\label{Fig10a}
}
			\subfigure[]{ 
	\includegraphics[scale=0.55]{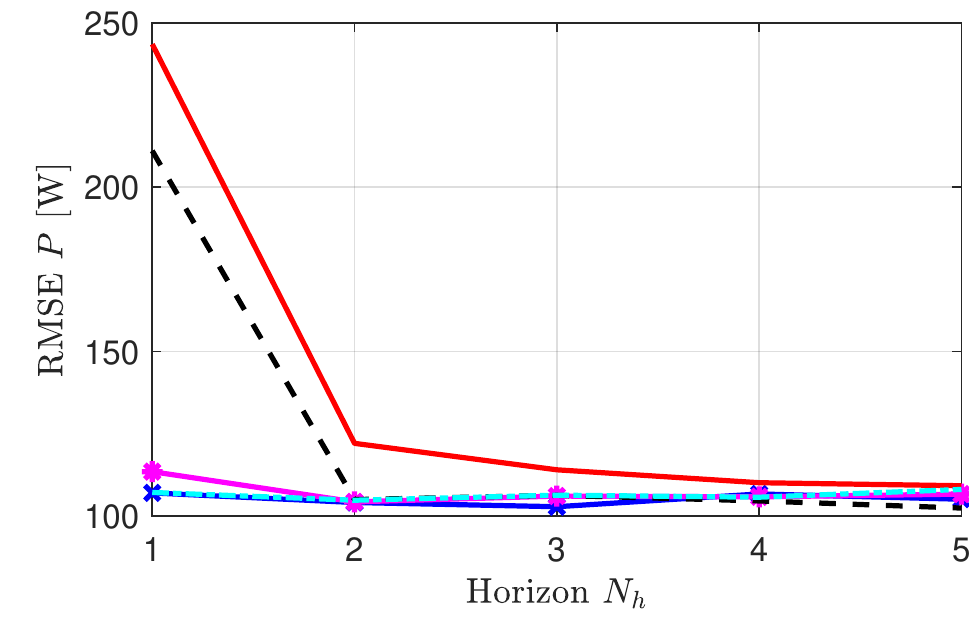}
	\label{Fig10b}
}
			\subfigure[]{ 
	\includegraphics[scale=0.55]{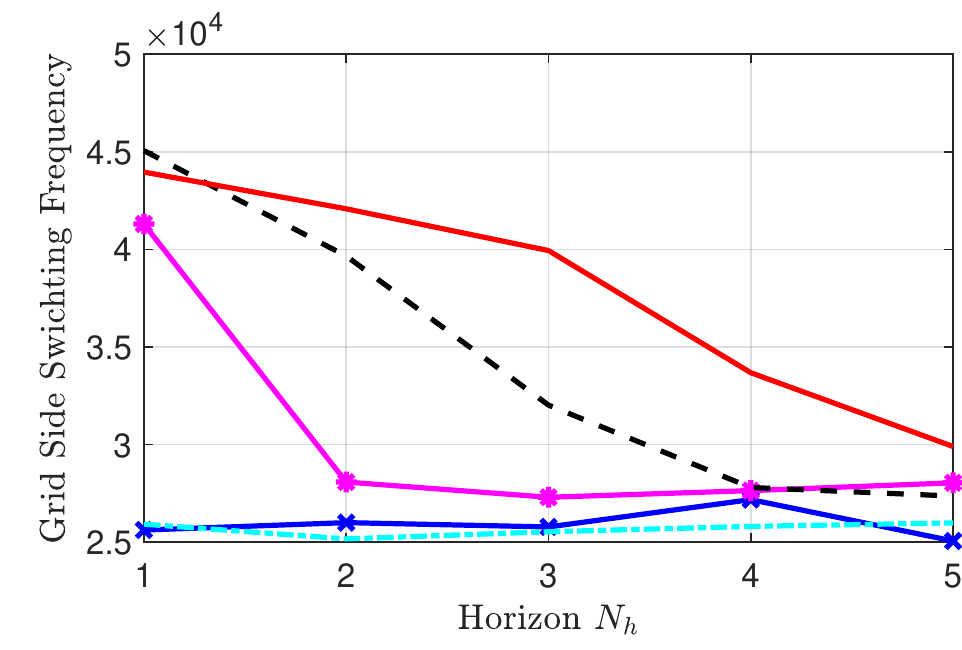}
	\label{Fig10c}
}
			\subfigure[]{ 
	\includegraphics[scale=0.55]{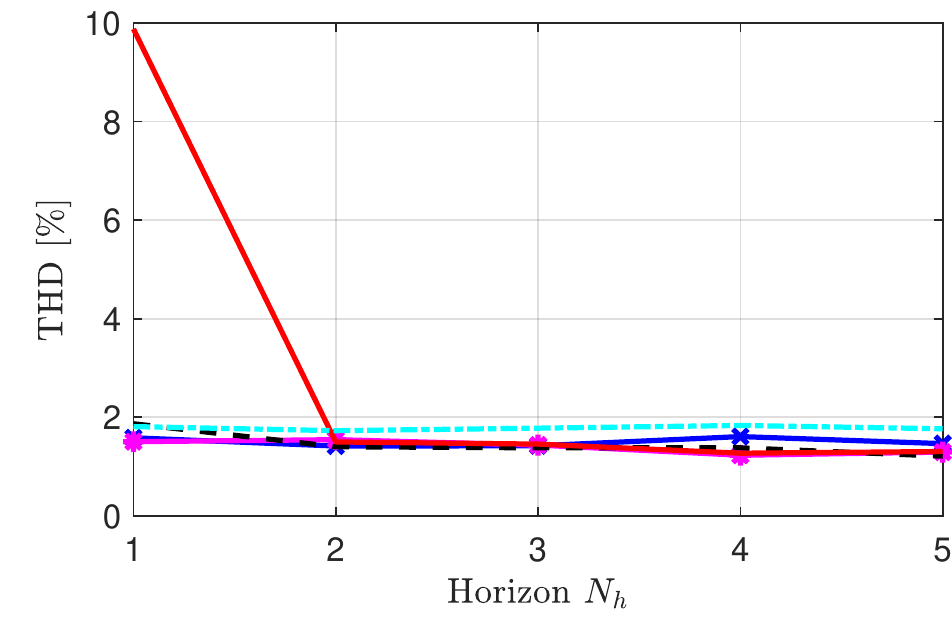}
	\label{Fig10d}
}
			\subfigure[]{ 
	\includegraphics[scale=0.55]{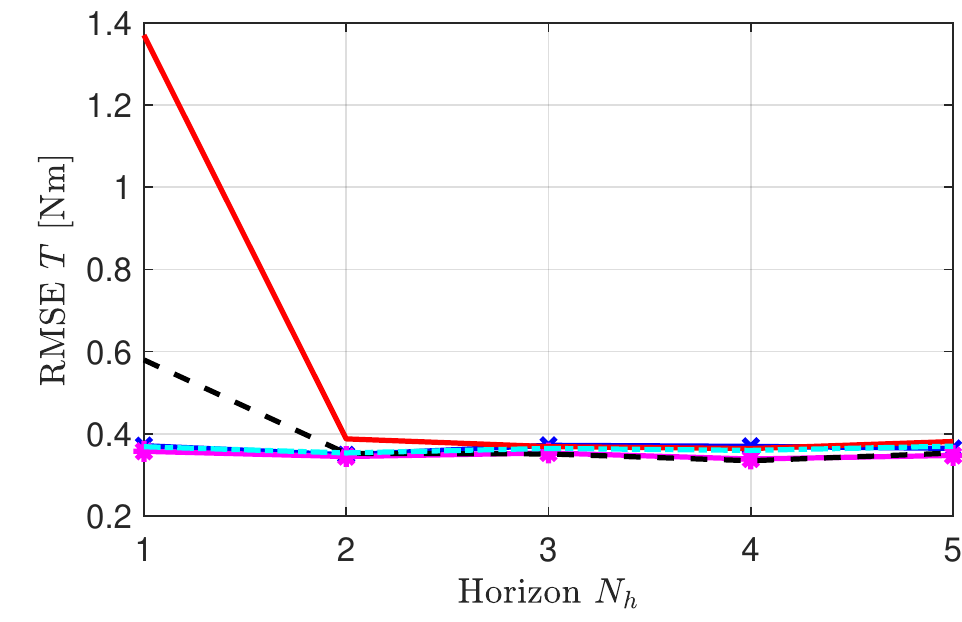}
	\label{Fig10e}
}
			\subfigure[]{ 
	\includegraphics[scale=0.55]{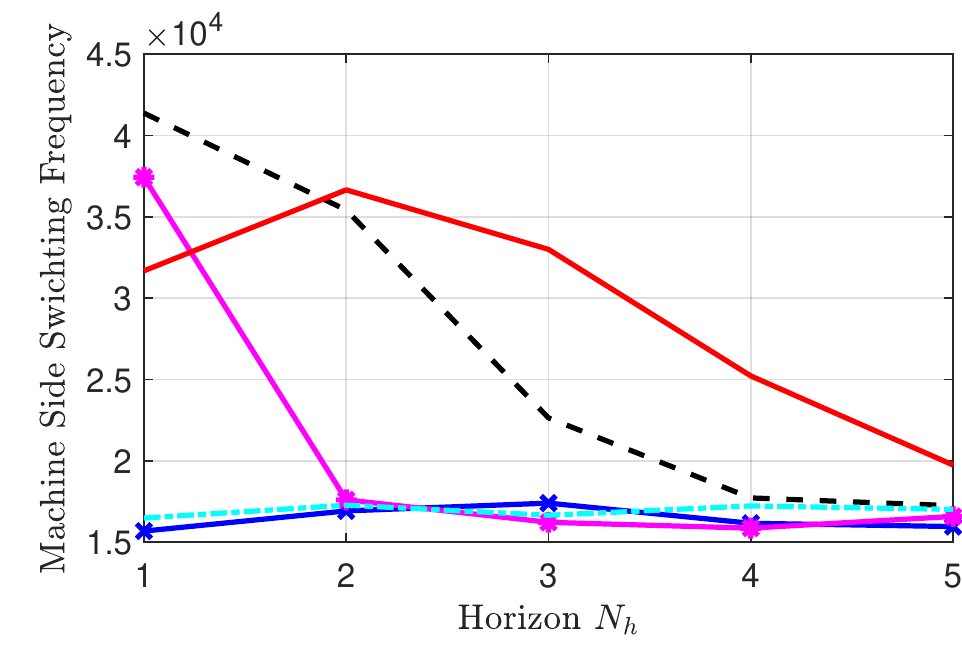}
	\label{Fig10f}
}
			\subfigure[]{ 
			\hspace{-0.9cm}
	\includegraphics[scale=0.55]{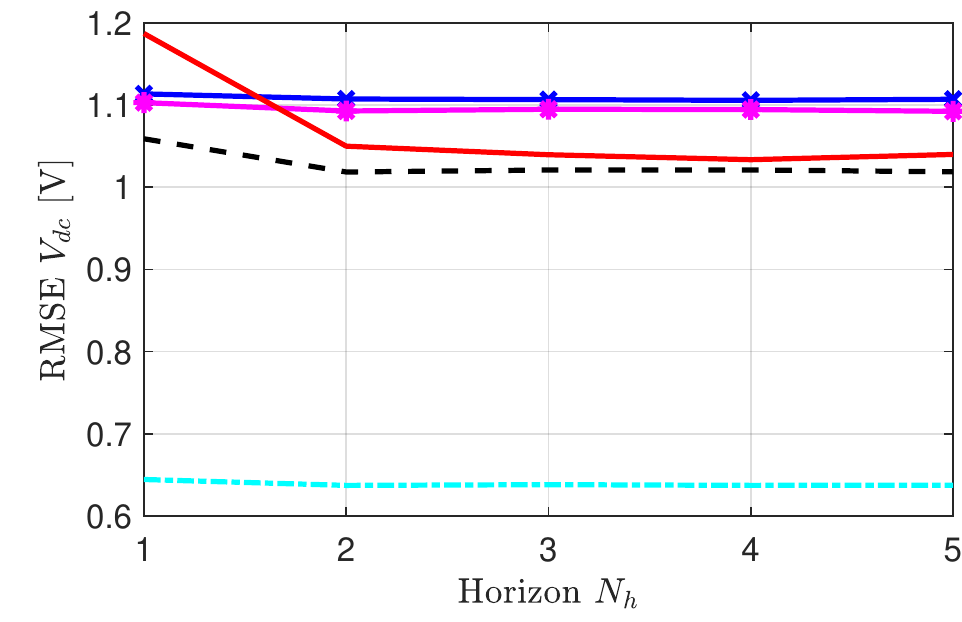}
	\label{Fig10g}
}
			\subfigure[]{ 
			\hspace{0.2cm}
	\includegraphics[scale=0.55]{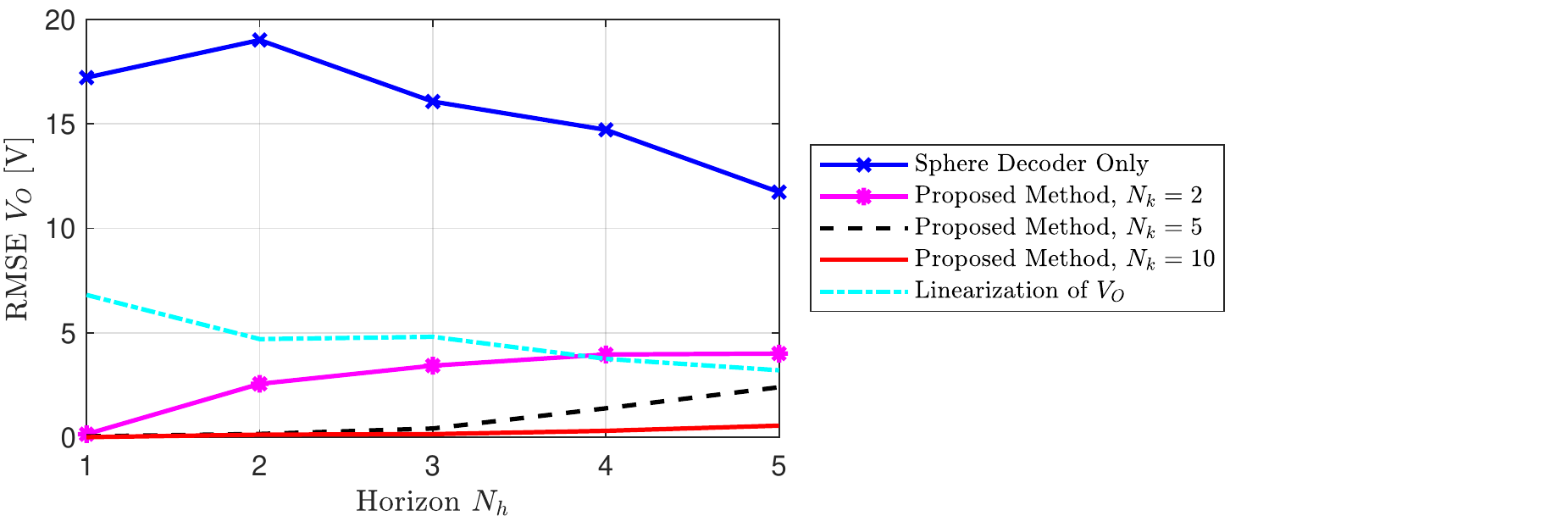}
	\label{Fig10h}
}
		\caption{Comparative analysis of different multistep MPC strategies. All algorithms were run on the case study system for different horizon lengths. The RMSE of the grid side reactive power $Q$ and reactive power $P$ is shown in a) and b). The comparison of the grid side switching frequencies can be seen in c). The machine side current THD and RMSE of the torque $T_e$ are shown in d) and e), and the machine side switching frequency in f).
	    The RMSE of the DC-link voltage $V_{dc}$ and DC-link balance $V_O$ are depicted in g) and h) respectively.
}
\label{fig:Comparison}
\end{center}
\end{figure*}
To prove the effectiveness of the proposed framework, it is compared to state of the art long horizon MPC methods. 
Therefore, the following algorithms were evaluated for different horizon lengths:
\begin{enumerate}
    \item A standard sphere decoding algorithm with no DC-link balancing, as presented in \cite{Multi}.
    \item The proposed method for $N_k=1$, $N_k =5$, and $N_k = 10$.
    \item A linearization of the DC-link balance in combination with the standard sphere decoder with $\lambda_V = 0.01$, as presented in \cite{FG3}.
\end{enumerate}
The results are shown in Fig. \ref{fig:Comparison}.
The grid side shows good performance for the proposed method for low $N_k$ if $N_h=1$, as shown in Fig. \ref{Fig10a} and Fig. \ref{Fig10b}. For higher values of $N_h$ such as e.g. $N_h = 5$ the proposed method improves and no difference is visible. The same can be observed for the grid side switching frequency which is shown in Fig. \ref{Fig10c}. For the machine side current THD and torque RMSE the effect is even stronger, (see Fig. \ref{Fig10d} and \ref{Fig10e}). For $N_h =2$ the 
RMSE of the torque which is related to the dynamic performance of the wind turbine, already shows similar performance to the standard sphere decoder and the linearization based approach.
The machine side switching frequency meanwhile is similar to the grid side switching frequency (see Fig. \ref{Fig10f}). 
For the DC-link voltage, there seems to be an optimal value of $N_k=5$ which always yields lower error compared to other values of $N_k$ (see Fig. \ref{Fig10g}). 
This shows that low values of $N_k$, the solution of the DC-link voltage is highly dependent on the optimal solution of the first optimization problem.
Regarding the performance of $V_O$, for higher $N_h$ the proposed algorithm reaches similar performance with state of the art algorithms for all figures of merit while still outperforming existing methods for the DC-link balance(see Fig. \ref{Fig10h}).
With an increasing number of the $N_k$ solution candidates that are used for nonlinear optimization, the likelihood of passing on those that yield bad performance for the system quantities increases as well. 
With an increasing horizon length however, the relation between $N_k$ and the total number of solution candidates decreases exponentially which explains the sharp increase in performance for $N_k=5$ and $N_k = 10$.
It can be seen from all Fig. except for Fig. \ref{Fig10h} that choosing large numbers of $N_k$ has a negative effect on the performance of the grid side and machine side figures of merit for shorter horizons. 
The decrease in switching frequency can be explained similarly as it is represented by the control effort which is minimized in the first optimization problem as well.
With increasing horizon length all system quantities quickly converge towards the state of the art methods.
The opposite can be seen for the nonlinear subsystem, i.e. the DC-link balance. 
In this case, the proposed algorithm outperforms both state of the art methods for $N_k >1$.
This indicates that sequential model predictive control is especially good for the nonlinear subsystems which are part of the second optimization stage.
Furthermore, in the case of multistep prediction, the performance for the linear subsystem quickly converges to the same value as the existing methods for all parameters. 
\section{Conclusion}\label{sec:conclusion}
We presented a long horizon cascaded model predictive approach for the control of Three-level NPC-PMSG wind turbine systems.
Therefore we split the system into two linear and one nonlinear subsystems where 
the cost functions for the linear subsystems are in quadratic form.
First, the minimizations of the quadratic forms were computed. 
Those solutions were then passed on to the nonlinear subsystems. 
The simulation results showed that the amount of solutions needed to achieve acceptable performance within the nonlinear subsystem is very small even for longer horizons.

\bibliographystyle{IEEEtran}
\bibliography{Paper}
%
%
%
\end{document}